\documentclass{article}

\RequirePackage{amsthm,amsmath,amsfonts,amssymb,mathrsfs,multirow}
\RequirePackage{graphicx}
\usepackage{natbib}

\newcommand{\bs}{\boldsymbol}

\begin{document}

\title{A Generalization of Ripley's K Function for the Detection of Spatial Clustering in Areal Data}

\author{Stella Self\textsuperscript{1*}, Anna Overby\textsuperscript{2}, Anja Zgodic\textsuperscript{1}, David White\textsuperscript{3}, \\
Alexander McLain\textsuperscript{1**}, Caitlin Dyckman\textsuperscript{2**}}

\maketitle

\noindent *Corresponding Author, scwatson@mailbox.sc.edu \\
**Shared Last Author \\
\textsuperscript{1}Arnold School of Public Health, University of South Carolina, 921 Assembly Street, Columbia, SC 29208, USA\\
\textsuperscript{2}College of Architecture, Arts and Humanities, Clemson University, Fernow Street, Clemson, SC 29634, USA\\
\textsuperscript{3}College of Behavioral, Social and Health Sciences, Clemson University, Epsilon Zeta Dr, Clemson, SC 29634, USA\\

\section*{Abstract}
Spatial clustering detection has a variety of applications in diverse fields, including identifying infectious disease outbreaks, assessing land use patterns, pinpointing crime hotspots, and identifying clusters of neurons in brain imaging applications.  While performing spatial clustering analysis on point process data is common, applications to areal data are frequently of interest.  For example, researchers might wish to know if census tracts with a case of a rare medical condition or an outbreak of an infectious disease tend to cluster together spatially.  Since few spatial clustering methods are designed for areal data, researchers often reduce the areal data to point process data (e.g., using the centroid of each areal unit) and apply methods designed for point process data, such as Ripley's K function or the average nearest neighbor method.  However, since these methods were not designed for areal data, a number of issues can arise. For example, we show that they can result in loss of power and/or a significantly inflated type I error rate. To address these issues, we propose a generalization of Ripley's K function designed specifically to detect spatial clustering in areal data.  We compare its performance to that of the traditional Ripley's K function, the average nearest neighbor method, and the spatial scan statistic with an extensive simulation study. We then evaluate the real world performance of the method by using it to detect spatial clustering in land parcels containing conservation easements and US counties with high pediatric overweight/obesity rates.



\section{Introduction}

The rapid growth in the use of geographic information system (GIS) software over the past thirty years has led to an explosion of spatial data and associated analytical methods.  Spatial data generally falls into one of two categories.  Point process data are associated with a specific latitude-longitude location, while areal data are associated with a spatial region (such as a county or census tract).  The locations of trees in a forest, the addresses of cases of an infectious disease, and locations of violent crimes are all examples of point process data.  Researchers frequently wish to determine if such data exhibit spatial clustering, loosely defined as an excess of events in one or more areas.  Areal data can also exhibit clustering.  Census blocks having a case of a rare cancer, land parcels with development restrictions, or counties which required individuals to wear a mask in public indoor settings during the COVID-19 pandemic are all examples of areal data which could be clustered.  While a variety of methods have been developed to assess point process data for clustering, less attention has been paid to the areal case.  

In this paper, we consider the problem of assessing binary areal data for spatial clustering and dispersion.  We stress that we are here considering clustering  in areal data based on location only, that is, we are attempting to answer the question, `is there an excess of areal units having some binary characteristic of interest in certain part(s) of the study area?'.  The terms `clustered data' or `clustering' are sometimes use to describe data which exhibits positive spatial autocorrelation in some numeric attribute.  For example, census tracts with a high incidence of an infectious disease might tend to be closer to other tracts with high incidence.  A number of methods exist for assessing areal data for this sort of clustering, including the Getis Ord Gi* statistic and the local Moran's I statistic, but `clustering' in the sense of positive spatial autocorrelation among data attributes is not the focus of this work.  For our purposes, the term `clustered data' or `clustering' will refer clustering based on location only, that is, we will say data is \emph{clustered} if the \emph{locations} of the observations are clustered, regardless of any spatial patterns present in their other attributes. 

Most cluster detection methods are either designed for point process data or intended to assess spatial autocorrelation in numeric data attributes derived from aggregating data over an areal unit.  One notable exception is the spatial scan statistic, initially developed in \cite{Kulldorff97}. Under the original formulation, the number of observations in the study area was assumed to follow either a binomial or a Poisson distribution.  The method identifies the `most likely cluster' by considering a large number of possible zones (usually circles) and performing a series of likelihood ratio tests to determine if probability of observing an event (under the binomial likelihood) or the event rate (under the Poisson likelihood) is larger inside the zone than outside.  The spatial scan statistic may be used to assess areal data for the presence of clustering by assuming a binomial likelihood where each areal unit has a population of 1, with units having the characteristic of interest considered to have 1 `success' \citep{Kulldorff06}.  When the spatial scan statistic is applied in this way, areal units are considered part of a zone if their centroid falls in the zone.  However, representing an areal unit with its centroid is not without problems.  If an areal unit is not convex, the centroid may lie outside the unit.  Two large adjacent units might have centroids which are quite distant from each other, even though the units themselves share a border. The spatial scan statistic method has been extended beyond the original binomial and Poisson cases to handle normally distributed data \citep{Huang09, Shen14}, time-to-event data \citep{Huang07, Bhatt16}, and ordinal data \citep{Jung07}; a non-parametric version has also be developed \citep{Matos21}.  \cite{Kulldorff06} develop a more flexible spatial scan statistic based on elliptical zones, while \cite{Tango2005} use irregularly shaped windows built from the underlying areal units. Typically Monte Carlo simulations are used to simulate the null distribution of the spatial scan statistic, though exact or asymptomatic null distributions are available for a few related variants \citep{Soltani15}.

One of the earliest methods for the detection of spatial clustering was the average nearest neighbor (ANN) method developed in \cite{Clark54}.  This method was developed for point process data.  The ANN method computes the average distance between each observed point and the point closest to it and compares this average to the expected distance under a null hypothesis of complete spatial randomness.  Small ANN ratio values indicate spatial clustering, while large values indicate dispersion.  Several extensions to the original ANN approach have been proposed \citep{Clark55, Clark56, Clark79}, and the method has been widely used in a variety of disciplines, including geography and ecology;  for a nice overview, see \cite{Philo21}.  While the ANN method was developed for point process data, it is commonly applied to areal data by using the centroids of areal units as the observed latitude and longitude locations.  ArcGIS software does this by default when applying the method to areal (polygon) data \citep{ArcANN}.  In such scenarios, the centroid of an areal unit is considered an `observed point' if the areal unit has some binary characteristic of interest.

Spatial clustering can occur at various geographical scales.  For example, cases of an infectious disease may be clustered within a household while infected households themselves are clustered at the neighborhood level.  Ripley's K function is a popular method for assessing spatial clustering because it allows researcher to assess clustering at a specific geographical scale (or multiple scales) \citep{Ripley76, Ripley77, Ripley81}.  For a given distance $r$, Ripley's K function provides a means of comparing the number of pairs of observations located within a distance of $r$ of each other and the number of such observations we would expect to find if the data were randomly scattered (i.e. no clustering).  If the overall density of observations in the space is $\lambda$ and there is no clustering, then we would expect a circle of radius $r$ centered at any given observation to contain approximately $\pi r^2 \lambda$ observations.  Values much larger than this indicate spatial clustering, i.e. the number of nearby points is larger than expected, while values much smaller than this are indicative of spatial dispersion.  Ripley's K function has been widely used in a variety of fields, including ecology \citep{Haase95}, microbiology \citep{Yunta14}, cancer detection \citep{Martins09}, image analysis \citep{Amgad15}, and archaeology \citep{Sayer13}.  After the initial development of Ripley's K function in Ripley's seminal works \citep{Ripley76, Ripley77, Ripley81}, a variety of extensions and adjustments were developed. Many of these adjustments are designed to address the problem of \emph{edge effects} caused by the underestimation of $K(r)$ near the study area boundary if some of the points within distance $r$ of a point in question fall outside of the study area.  Edge corrections for circular \citep{Diggle83}, rectangular \citep{Diggle83} and irregular \citep{Goreaud99} study areas have been developed, as well as several more complex methods of edge correction \citep{Sterner86, Szwagrzyk93, Upton85, Getis87, Andersen92}. 

Inference based on Ripley's K function rests heavily on the theory of spatial point processes. For example, the null hypothesis for statistical tests based on Ripley's K function is that the points arise from a two dimensional homogeneous Poisson process, which is inherently violated by areal data. However, the lack of cluster detection methods designed specifically for areal data have caused many researchers to use Ripley's K function on areal data, typically using the centroids and computing Ripley's K function using the resulting set of points.  For example, a number of researchers have attempted to assess spatial patterns in land parcel data via Ripley's K function or Ripley's L function (a scaled version of Ripley's K function) \citep{Lee13, Siordia13, Zipp17,Qiao19}.  Other researchers have taken public health data associated with a geographical region such as a city or health division and computed Ripley's K function using the centroids of these larger geographical regions \citep{Wade14, Karunaweera2020, Skog2014}.  Ripley's K function has also been used to assess areal data for clustering in a variety of ecological and geological applications \citep{Kretser, davarpanah2018spatial,Tonini13}.  In fact, ArcGIS software computes Ripley's K function for areal (polygon feature) data by mapping each areal unit to its centroid by default \citep{arcRipK}.  To our knowledge, the performance of Ripley's K function on areal data has never been evaluated. We show in our simulation studies that Ripley's K function often has a severely inflated type I error rate when applied to areal data.  

In this paper, we propose an extension of Ripley's K function, which we refer to as Ripley's K function for areal data (RKAD). The interpretation of RKAD is similar to that of the traditional Ripley's K function, but it possesses improved performance for areal data.  In Section 2, we define Ripley's K function and RKAD.  Section 3 presents the results of an extensive simulation study to compare performance of RKAD, Ripley's K function, the ANN method, and the spatial scan statistic when assessing areal data for clustering or dispersion.  In Section 4, we assess the performance of RKAD on two real datasets.  First, we use it to determine if land parcels in Boulder County, Colorado which contain a conservation easement are spatially clustered.  Next, we apply RKAD to determine if US counties with high childhood overweight rates are spatially clustered.  Section 5 provides concluding remarks.


\section{Methodology}\label{sec2}

\subsection{Ripley's K Function}

Suppose we are observing a spatial point process on a two dimensional region $\mathscr A$, with density $\lambda$.  For a distance $r>0$, Ripley's $K$ function is defined as 
\[
K(r) = \lambda^{-1}E(\text{number of points within a distance of $r$ of any given point}).
\]
If we have observed a collection of $n$ points, we can estimate Ripley's K function as
\[
\hat K(r) = \hat \lambda^{-1} \sum_{i = 1}^n \sum_{j = 1}^n \frac{w_{ij}}{n}
\]
where $\hat \lambda = n/|\mathscr A|$, $|\mathscr A|$ denotes the area of $\mathscr A$, and $w_{ij}$ is a weight associated with points $i$ and $j$.  In the traditional approach, $w_{ij} = 1$ if the distance between points $i$ and $j$ is less than $r$ and 0 otherwise \citep{Ripley76, Ripley77}.  However, many variants of Ripley's K function exists which modify these weights to account for edge effects \citep{Ripley76,Sterner86,Diggle83, Szwagrzyk93, Upton85, Getis87, Andersen92}.  In practice, Ripley's K function is often re-scaled to Ripley's L function, $\hat L(r) = \{\hat K(r)/\pi\}^{1/2}$.

Ripley's K function is often used to determine if an observed collection of points exhibits complete spatial randomness (CSR) (i.e. the points arise from a two-dimensional homogeneous Poisson process).  For a homogeneous Poisson process on an infinite study area, $K(r) = \pi r^2$.  The distribution of $\hat K(r)$ under CSR for a finite study area $\mathscr A$ can be approximated with Monte Carlo simulations, which are used to perform a hypothesis test with the null hypothesis being that the observed data arises from a homogeneous Poisson process with rate parameter $\hat \lambda$.  Large values of $\hat K(r)$ indicate spatial clustering, that is, the number of points within a distance of $r$ of any given point is larger than would be expected if the data exhibited CSR.  Small value of $\hat K(r)$ indicate dispersion, that is, the number of points within a distance of $r$ of any given point is smaller than would be expected under CSR.  For an overview of Ripley's K function, edge correction methods, and hypothesis testing via Monte Carlo simulations, see \cite{Dixon14}.

\subsection{Ripley's K Function for Areal Data}

While Ripley's K function is designed for point process data, in practice, it is often used to assess areal data for clustering.  Areal units with some binary characteristic of interest are mapped to their centroids and Ripley's K function is applied to to the resulting set of points.  The lack of suitable alternative methods to assess clustering in areal data as well as the fact that ArcGIS applies Ripley's K to polygon centroids by default both contribute to this misuse \citep{arcRipK}. Applying Ripley's K to the centroids of areal units is particularly problematic when the units are vastly different sizes. For example, in one of our motivating data applications we wish to determine if land parcels having conservation easements (CEs) are clustered.  Under the null hypothesis, all parcels are equally likely to have a CE.  Under the null hypothesis, a portion of the study area with many small parcels (such as a metropolitan area), will have more parcels with CEs than portions of the study area with many large parcels, simply because there are more parcels per unit area.  Put another way, centroids of smaller parcels will appear clustered relative to centroids of larger parcels simply because the size of the small parcels allows the centroids to be closer together.  

\begin{figure}[b]
    \centering
    \includegraphics[scale = 0.3]{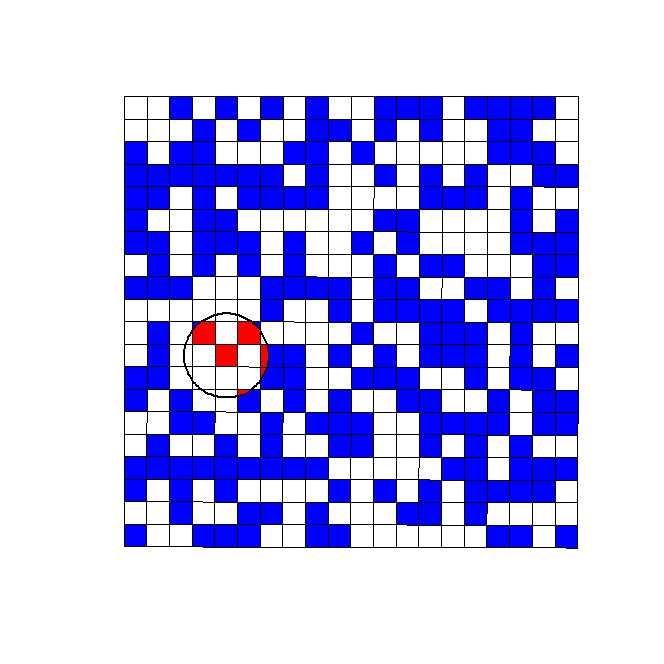}
    \caption{An illustration of $c_i(r) \cap \mathscr B(\bs y)$ for a randomly selected areal unit.  Observed areal units (i.e. units for which $y_i = 1$ are shown in color.  The red area corresponds to $c_i(r) \cap \mathscr B(\bs y)$ for a selected unit $a_i$ and radius $r$.}
    \label{fig:intersection}
\end{figure}

To surmount these difficulties, we propose a modification to Ripley's K function.  Suppose we have a study area $\mathscr A$ divided into $N$ areal units $a_1,a_2,...,a_N$, and that each unit $a_i$ possesses some characteristic of interest with probability $p_i$.  For each unit, we define the Bernoulli$(p_i)$ random variable $Y_i$ to be 1 if unit $i$ possesses the trait of interest and 0 otherwise. For example, $\mathscr A$ might be a county, $a_1$,...$a_N$ be the parcels of land in the county, and $Y_i = 1$ if parcel $i$ contains a CE.  We will refer to the subset of the units for which $Y_i = 1$ as the `observed units' or the `observations', and denote them by $\mathscr B(\bs Y) = \{ a_i : Y_i = 1 \}$, where $\bs Y = (Y_1,...,Y_N)'$.  For a given study area $\mathscr A$, set of areal units $a_1,...,a_N$, and observed data $\bs y$, for $r>0$ and each $i$ for which $ y_i = 1$, define
\[
m(r,i, \bs y) = \frac{|c_i(r) \cap \mathscr B(\bs y) |}{\pi r^2}
\]
where $c_i(r)$ denotes the circle of radius $r$ centered at the centroid of $a_i$.  Thus $m(r,i, \bs y)$ is the proportion of the circle of radius $r$ centered at the centroid of $a_i$ which falls into the observed units (see Figure \ref{fig:intersection}).  Note that $m(r,i,\bs y)$ is only defined if $a_i \in \mathscr B(\bs Y)$.  Additionally, define $m(r, \bs y)$ as the sample average (taken over all $n_{\bs y} = \sum_{i=1}^N y_i$ observed units) of the amount of area within a distance of $r$ of an observed unit centroid which falls into observed units.
\[
m(r,\bs y) = \frac{1}{n_{\bs y}}\sum_{i: y_i = 1}  m(r,i,\bs y).
\]

Finally, define Ripley's K function for areal data (RKAD) by
\[
m(r) = E \{ m(r,\bs Y)\}  = \sum_{\bs y \in \mathscr Y} \left( \frac{1}{n_{\bs y}}\sum_{i: y_i = 1}  \frac{|c_i(r) \cap \mathscr B(\bs y) |}{\pi r^2} \right)P(\bs Y = \bs y)
\]
where $\mathscr Y = \{ (y_1,...,y_N): y_i \in \{ 0,1 \} \}$. Therefore,
\[
m(r) = \pi^{-1} r^{-2}E(\text{observed area within a distance of $r$ of an observed unit centroid}).
\]
Also define
\[
m(r;n_{\bs y}) = E\{m(r,\bs Y)| n_{\bs Y} = n \} =  \sum_{\bs y \in \mathscr Y_n} \left( \frac{1}{n}\sum_{i: y_i = 1}  \frac{|c_i(r) \cap \mathscr B(\bs y) |}{\pi r^2} \right)P(\bs Y = \bs y|n_{\bs Y} = n)
\]
where $\mathscr Y_n = \{ \bs y \in \mathscr Y : \sum_{i=1}^N y_i = n \} $.
Thus
\begin{align*}
m(r;n) = \pi^{-1} r^{-2} E(&\text{observed area within a distance of $r$ of an observed unit centroid}\\
&|\text{n units were observed})
\end{align*}
While $m(r)$ and $m(r;n_{\bs y})$ may be calculated explicitly provided the vector of probabilities $\bs p = (p_1,...,p_N)'$ is known, doing so involves computing (respectively) a $2^N$ and an $\binom{N}{n_{\bs y}}$ dimensional sum, and Monte Carlo approximations are likely to be more practical when $N$ is large.

\subsection{A Test for Spatial Clustering or Dispersion Based on RKAD}

A test for clustering or dispersion in the observed units may be derived by comparing the observed $m(r,i,\bs y)$ values to $m(r; n_{\bs y})$ calculated in the absence of spatial dependence.  Consider the following test statistic
\[
T(r,\bs y) = \frac{1}{n_{\bs y}}\sum_{i: y_i = 1} |m(r,i, \bs y) - m(r;n_{\bs y})|
\]
where $m(r; n_{\bs y} )$ is calculated under an assumed null distribution which lacks spatial dependence. Values of $T(r,\bs y)$ which are large or small relative to the null distribution indicate a departure from the null distribution.  When units exhibit clustering, certain part(s) of the study area will contain more units than expected under the null distribution.  As a consequence, other parts of the study area will have fewer units than expected (as the areas in between the clusters must be sparse in comparison).  Therefore values of $m(r,i,\bs y)$ which are much larger or much smaller than $m(r; n_{\bs y})$ are \emph{both} indicative of clustering.  Consequently, a large value of $T(r, \bs y)$ (i.e. a large average absolute difference between the observed $m(r,i, \bs y)$'s and the estimated value $m(r, n_{\bs y})$ under the null distribution) indicates spatial clustering.  When the areal units are dispersed, that is, distributed at more regular intervals than expected in the absence of spatial dependence, then variability in the $m(r, i, \bs y)$'s will be decreased and the average absolute difference between the $m(r,i, \bs y)$s and $m(r,n_{\bs y})$ under the null distribution will be small.  Therefore, small values of $T(r, \bs y)$ indicate dispersion.  

While the sampling distribution of $T(r,\cdot) $ is not amenable to direct evaluation, it can be approximated via Monte Carlo simulations.  The same Monte Carlo simulations can be used to approximate $m(r;n_{\bs y})$.  Such an approximation involves specification of the null distribution.  The class of distributions which lack spatial dependence is too broad to be practical, and the class must be narrowed in some way.  As is typical for hypothesis tests for spatial dependence, we take the case of independent and identically distributed observations as our null distribution (for exact specifications, see Section 3). An approximately $\alpha$-level hypothesis test for clustering can be conducted by rejecting the null hypothesis if $T(r, \bs y)$ exceeds the $1-\alpha$th quantile of the Monte Carlo sample of $T(r, \cdot)$.  Similarly, a test for dispersion can be conducted by rejecting the null hypothesis if $T(r,\bs y)$ is less than the $\alpha$th quantile of the Monte Carlo sample.

\section{Simulation Study}\label{sec3}

\subsection{Simulation Specifications}

\begin{figure}[t]
\centering
\includegraphics[scale = 0.35,trim={3cm 4cm 3cm 3cm},clip]{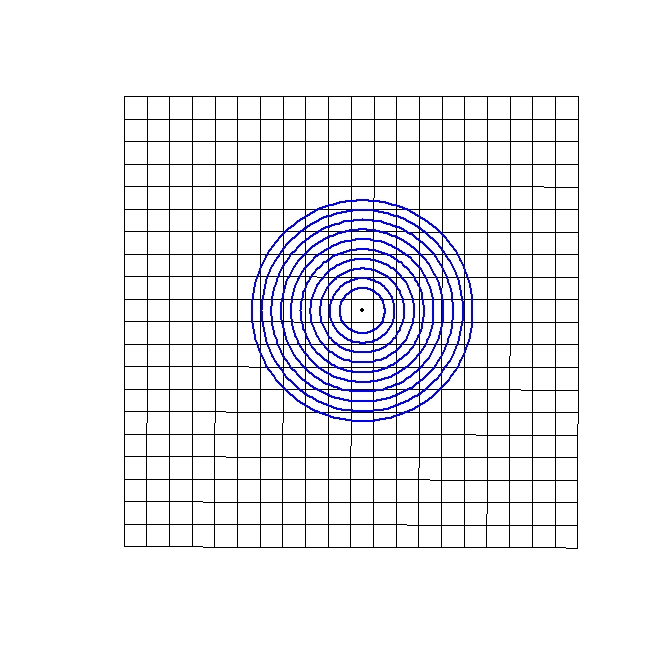}
\includegraphics[scale = 0.35,trim={3cm 0cm 2cm 4cm},clip,angle = 345]{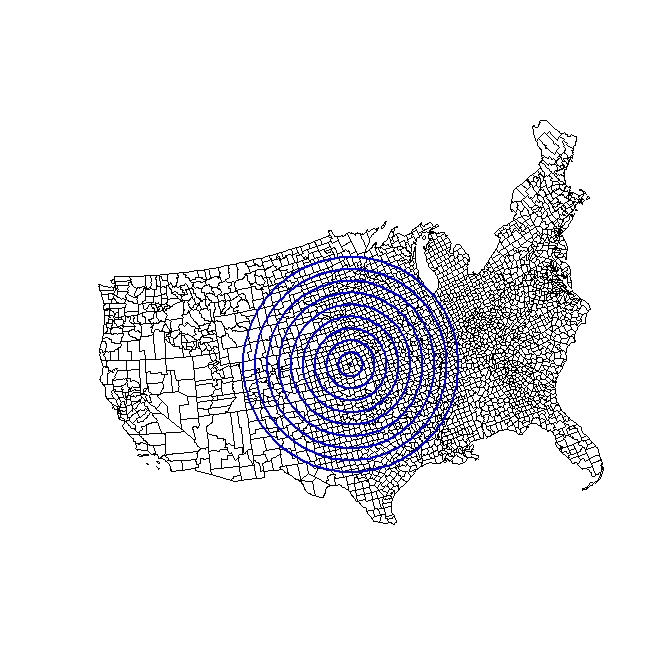}
\vspace{-0.5in}
\caption{The 2 study areas considered in the simulation study.  The 10 radii at which Ripley's K function and Ripley's K function for areal data are evaluated are shown for a single location in blue.\label{fig:simsunits}}
\end{figure}

In this section, we perform an extensive simulation study to compare the performance of RKAD to that of the ANN method, the spatial scan statistic, and the traditional Ripley's K function. We consider the performance of our proposed hypothesis testing procedure using two study areas which are shown in Figure \ref{fig:simsunits}:
\begin{align*}
    \mathscr A_1:& \text{ A 20 by 20 regular grid of $N_1 = 400$ cells}\\
    \mathscr A_2:& \text{ The $N_2 = 3,108$ counties (and county-equivalents) in the contiguous United States}
\end{align*}

\noindent For each study area $\mathscr A_j$, $j = 1,2$, we generate data under 19 different scenarios.  First we consider the case of no spatial pattern in the locations of the observed units via the following three scenarios:  
\begin{align*}
    \textbf{$I_1$: }& \bs Y \sim \text{SWoR}(N_j,\lceil N_j/10 \rceil,\bs p_{jI}) \\
    \textbf{$I_2$: }& \bs Y \sim \text{SWoR}(N_j,\lceil N_j/4 \rceil,\bs p_{jI}) \\
    \textbf{$I_3$: }& \bs Y \sim \text{SWoR}(N_j,\lceil N_j/2 \rceil,\bs p_{jI}).
\end{align*}
Here $\bs Y \sim $SWoR$(N,k,\bs p)$ indicates that the random variable $\bs Y = (Y_1,...,Y_N)$ arises by selecting $k$ elements from $\{1,2,..,N\}$ via sampling without replacement (SWoR) where $\bs p = (p_1,...,p_N)'$ gives the probability of selecting each element, and $Y_i = 1$ if $i$ was selected and 0 otherwise.  Areal unit $a_i$ is observed if and only if $y_i = 1$.  For $j = 1, 2$, we set $\bs p_{jI} = (N_j^{-1},\ldots, N_j^{-1})'$.   

\begin{figure}[t]
\[
\includegraphics[scale = 0.3,trim={3cm 4cm 3cm 3cm},clip]{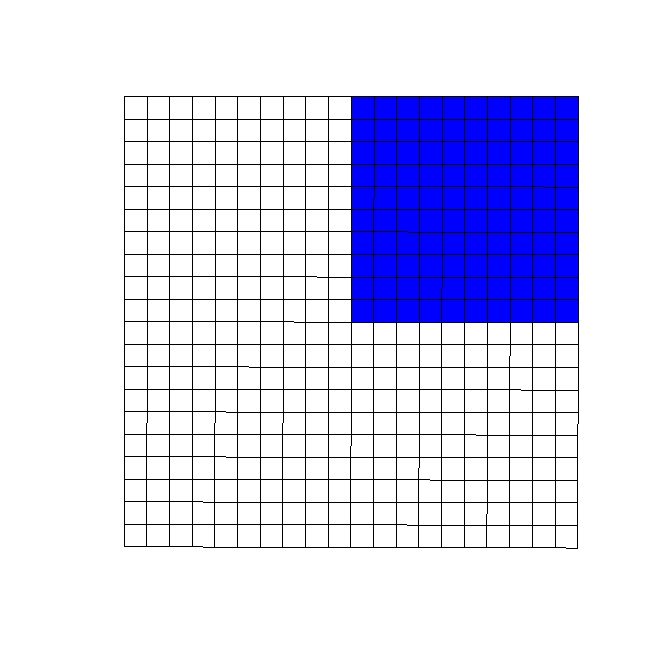} 
\includegraphics[scale = 0.35,trim={3cm 0cm 2cm 4cm},clip,angle = 345]{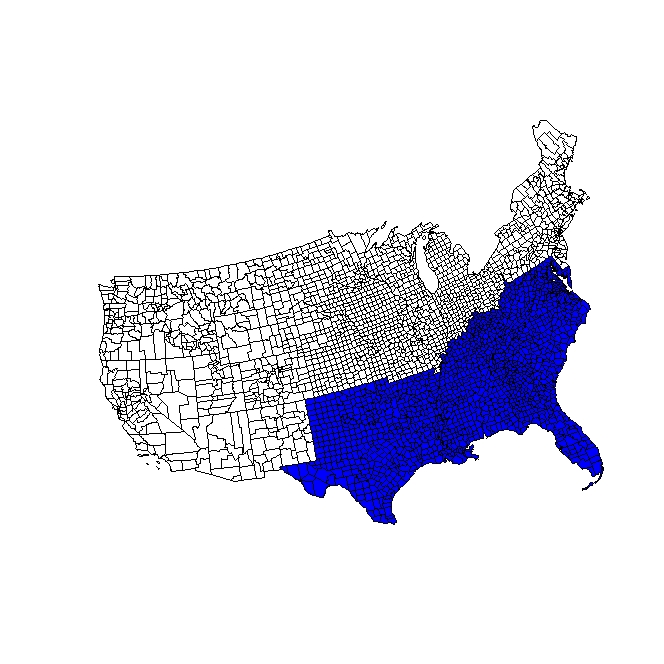}
\]

\vspace{-0.5in}

\caption{Illustration of spatial dependence structures $C_1-C_6$ for study areas $\mathscr A_1$ (left), and $\mathscr A_2$ (right). Blue units are $q$ times more likely to be selected than white units under spatial dependence configurations $C_1-C_6$.\label{fig:cluster}}
\end{figure}

For each study area, we also consider 12 scenarios in which the locations of observed units are clustered.  First, we assess the ability of our hypothesis test to detect large-scale spatial clustering which exists in only one part of the study area.  We consider the following 6 scenarios:
\begin{align*}
    \textbf{$C_1$: }& \bs Y \sim \text{SWoR}(N_j,\lceil N_j/10 \rceil,\bs p_{jC_1})\\
    \textbf{$C_2$: }& \bs Y \sim \text{SWoR}(N_j,\lceil N_j/10 \rceil,\bs p_{jC_2})\\
    \textbf{$C_3$: }& \bs Y \sim \text{SWoR}(N_j,\lceil N_j/4 \rceil,\bs p_{jC_3})\\
    \textbf{$C_4$: }& \bs Y \sim \text{SWoR}(N_j,\lceil N_j/4 \rceil,\bs p_{jC_4})\\
    \textbf{$C_5$: }& \bs Y \sim \text{SWoR}(N_j,\lceil N_j/2 \rceil,\bs p_{jC_5})\\
    \textbf{$C_6$: }& \bs Y \sim \text{SWoR}(N_j,\lceil N_j/2 \rceil,\bs p_{jC_6}).
\end{align*}
The $N_j$-dimensional vectors $\bs p_{jC_l} = (p_{jC_l1},\ldots,p_{jC_lN_j})'$, $j =1,2$, $l = 1,...,6$ are defined as follows: an entry of $p_{jC_l}$ is equal to $q/D$ if the unit is shown in blue in Figure \ref{fig:cluster} and equal to $1/D$ otherwise where $D$ is such that $\sum_i p_{jC_li}=1$.  For $C_1$, $C_3$ and $C_5$, we take $q = 5$ and for $C_2$, $C_4$ and $C_6$ we take $q = 10$.  Thus the blue units are 5 times more likely to be observed than the white units under data generation mechanisms $C_1$, $C_3$ and $C_5$ and 10 times more likely to be observed under data generation mechanisms $C_2$, $C_4$ and $C_6$.

Next, we assess the ability of our hypothesis test to detect spatial clustering at a smaller scale when it occurs at multiple clusters spread through the study area by considering the following 6 scenarios:
\begin{align*}
    \textbf{$C_7$: }& \bs Y \sim C(\lceil N_j/10 \rceil,\lceil N_j/100 \rceil,  5)\\
    \textbf{$C_8$: }& \bs Y \sim C(\lceil N_j/10 \rceil,\lceil N_j/100 \rceil,10)\\
    \textbf{$C_9$: }& \bs Y \sim C(\lceil N_j/4 \rceil,\lceil N_j/40 \rceil,5)\\
    \textbf{$C_{10}$: }& \bs Y \sim C(\lceil N_j/4 \rceil,\lceil N_j/40 \rceil, 10)\\
    \textbf{$C_{11}$: }& \bs Y \sim C(\lceil N_j/2 \rceil,\lceil N_j/20 \rceil,5)\\
    \textbf{$C_{12}$: }& \bs Y \sim C(\lceil N_j/2 \rceil,\lceil N_j/20 \rceil,10)
\end{align*}
Here, $\bs Y\sim C(k,m,q) $ denotes that $\bs Y $ is generated as follows.  First, $m$ elements of $\{1,2,...,N \}$ are randomly selected via SWoR$(N_j,m,p_{jI})$.  Next, $k-m$ elements of $\{1,2,...,N \}$ are selected via SWoR$(N,k-m,\bs p_2)$, where $p_{2i}= 0$ if $i$ was one of the first $m$ elements selected, $p_{2i} = q/D$ for $i$ such that $a_i$ shares a border with at least one of the initially selected $k$ elements (but $i$ itself was not initially selected) and $p_{2i} = 1/D$ otherwise, where $D$ is chosen so that $\sum_{i}p_{2i} = 1$; $\bs Y_i =1$ if element $i$ was selected in either the first or second step, and $\bs Y_i = 0$ otherwise, and areal unit $a_i$ is observed if and only $Y_i = 1$.  Note that data generation mechanisms $C_7$, $C_9$ and $C_{11}$ correspond to `weaker' clustering, in the sense that they tend to select fewer adjacent units than mechanisms $C_8$, $C_{10}$ and $C_{12}$

\begin{figure}[b]
\centerline{\includegraphics[scale = 0.11,trim={3cm 3cm 3cm 3cm},clip]{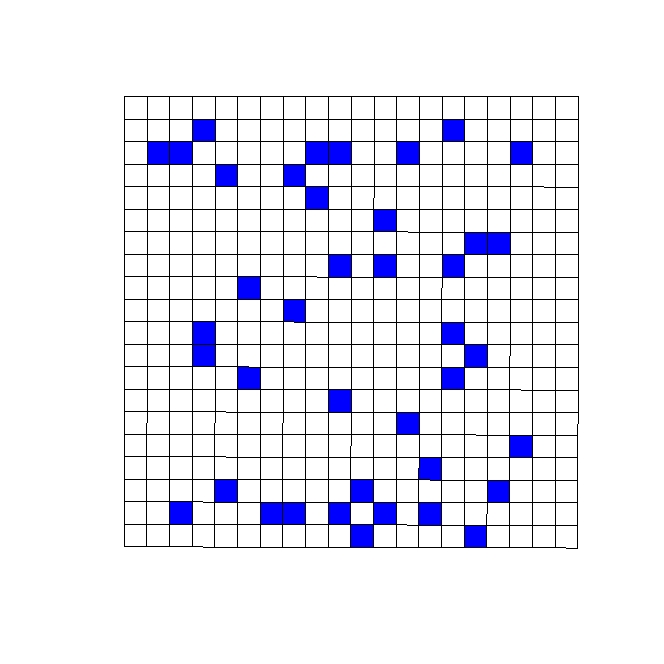}
\includegraphics[scale = 0.11,trim={3cm 3cm 3cm 3cm},clip]{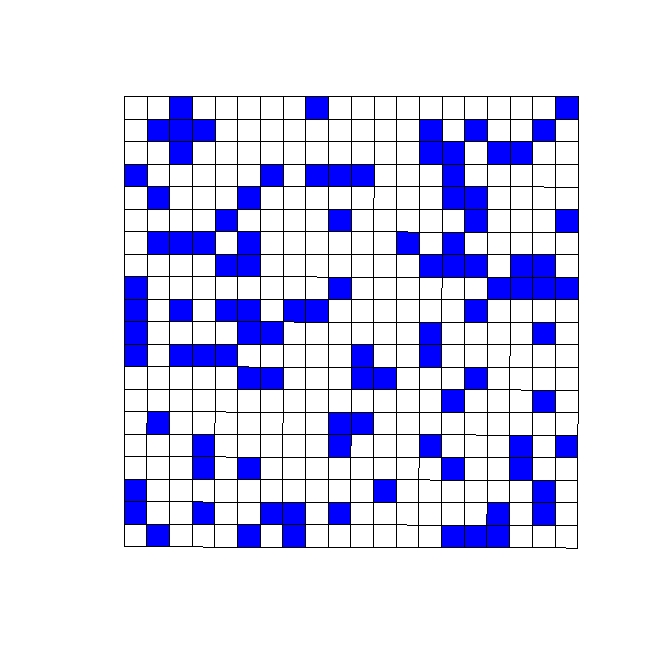}
\includegraphics[scale = 0.11,trim={3cm 3cm 3cm 3cm},clip]{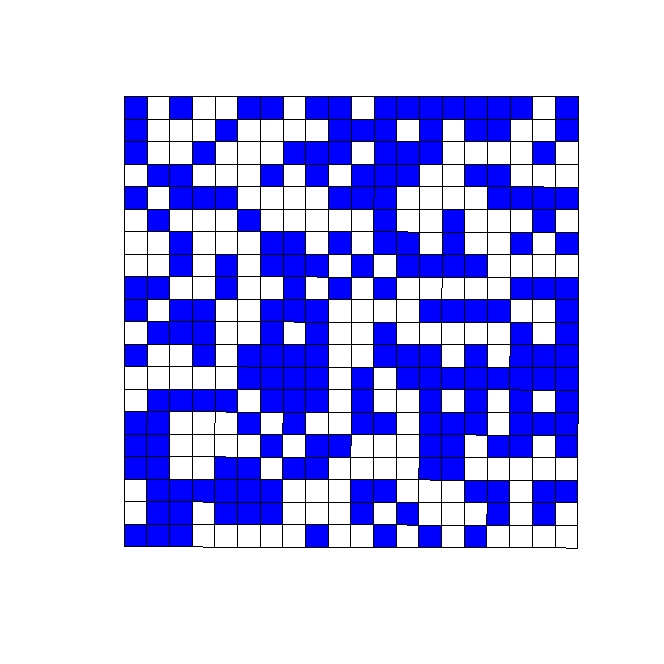}
}
\centerline{
\includegraphics[scale = 0.11,trim={3cm 3cm 3cm 3cm},clip]{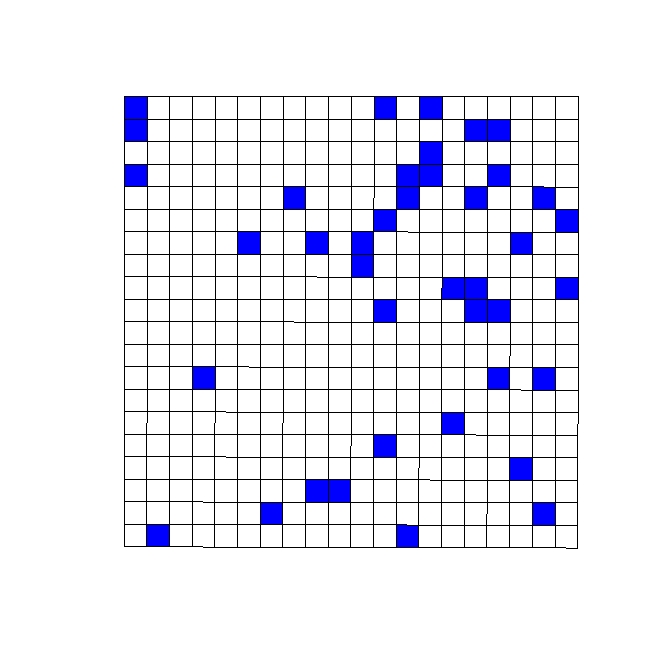}
\includegraphics[scale = 0.11,trim={3cm 3cm 3cm 3cm},clip]{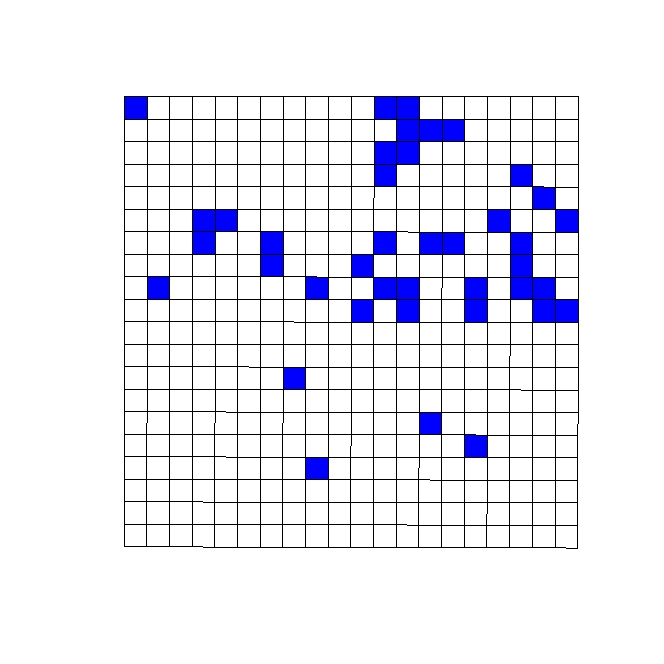}
\includegraphics[scale = 0.11,trim={3cm 3cm 3cm 3cm},clip]{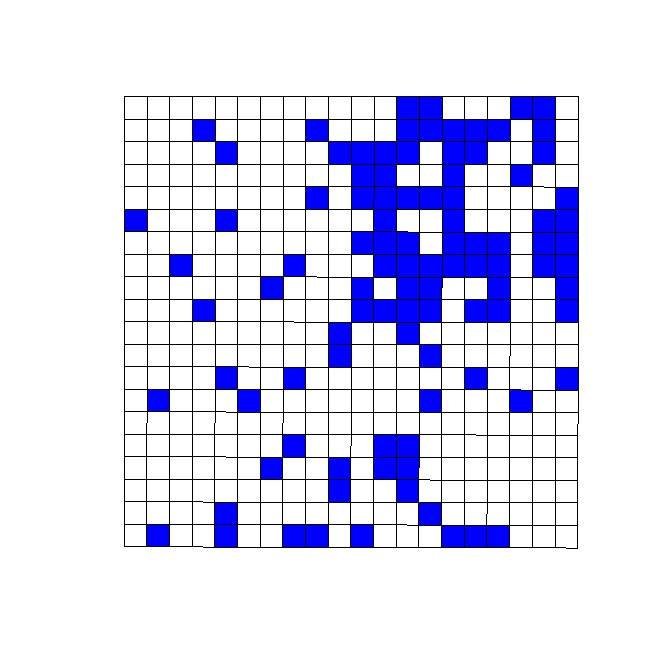}
\includegraphics[scale = 0.11,trim={3cm 3cm 3cm 3cm},clip]{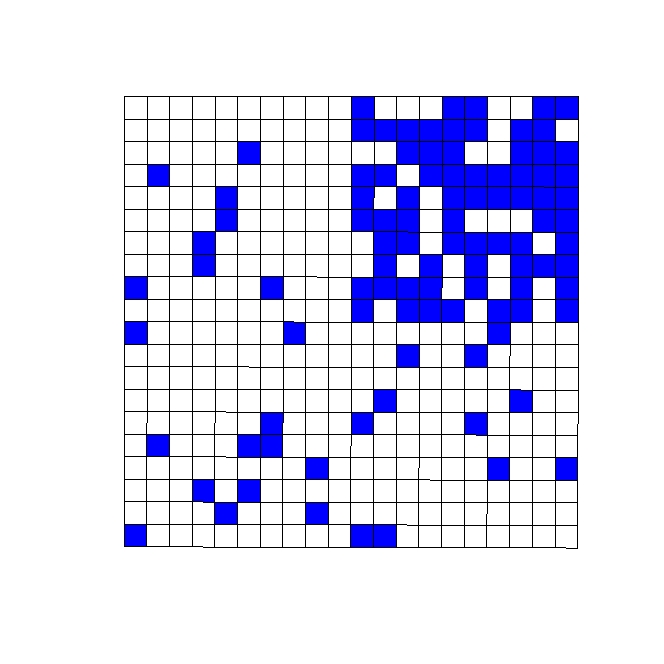}
\includegraphics[scale = 0.11,trim={3cm 3cm 3cm 3cm},clip]{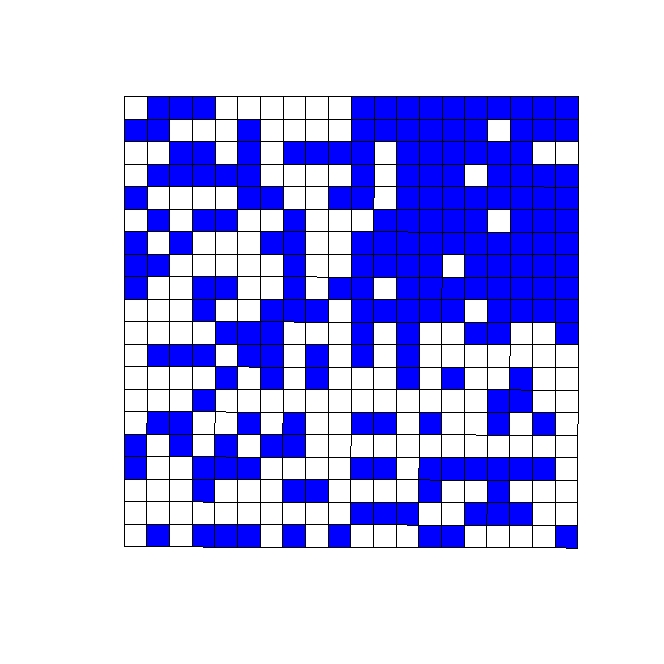}
\includegraphics[scale = 0.11,trim={3cm 3cm 3cm 3cm},clip]{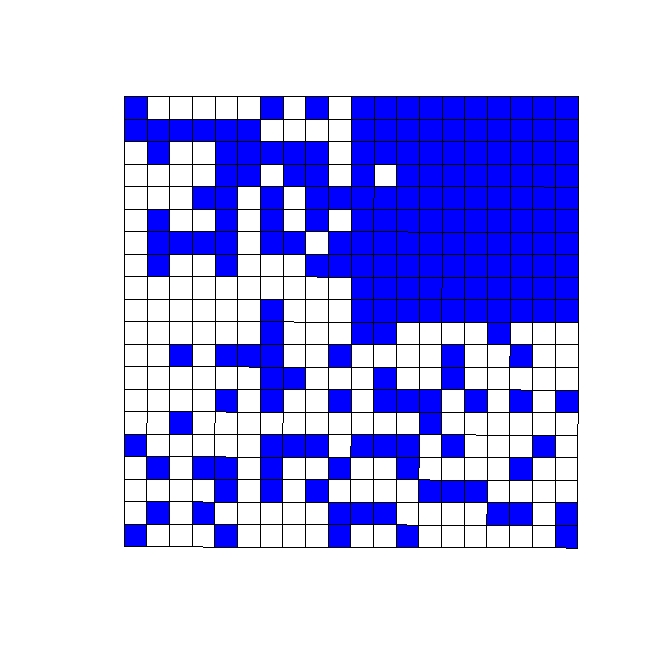}
}
\centerline{
\includegraphics[scale = 0.11,trim={3cm 3cm 3cm 3cm},clip]{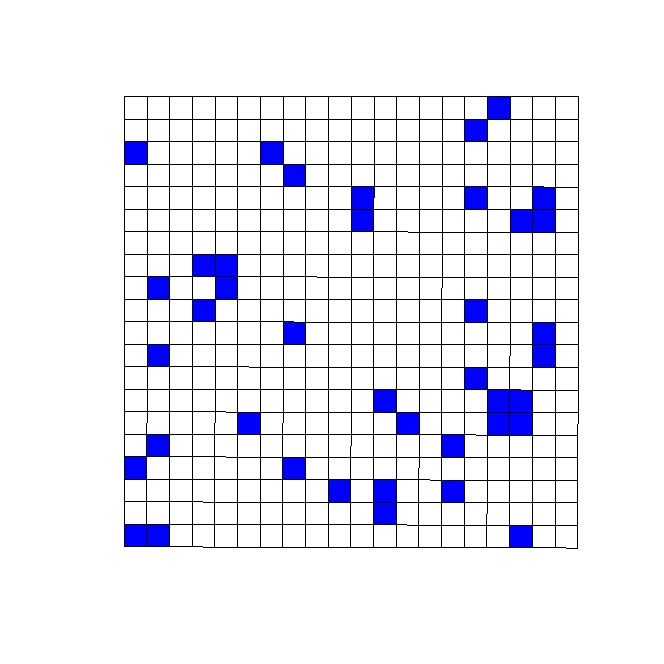}
\includegraphics[scale = 0.11,trim={3cm 3cm 3cm 3cm},clip]{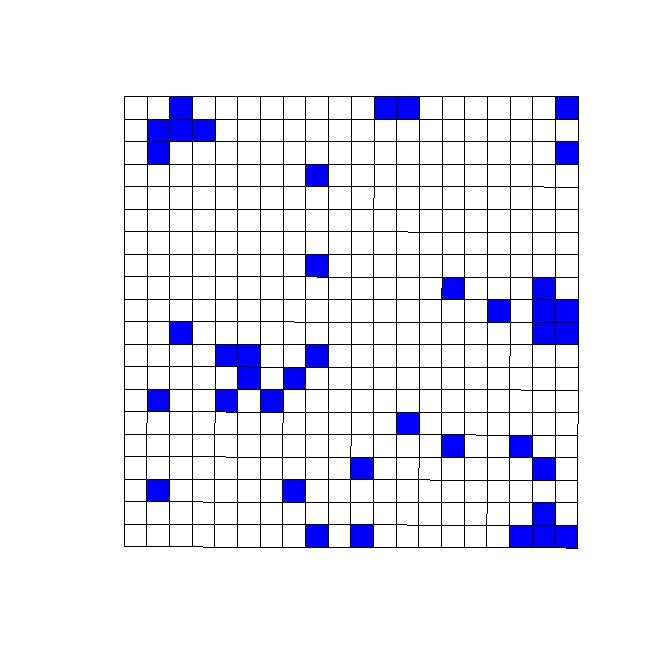}
\includegraphics[scale = 0.11,trim={3cm 3cm 3cm 3cm},clip]{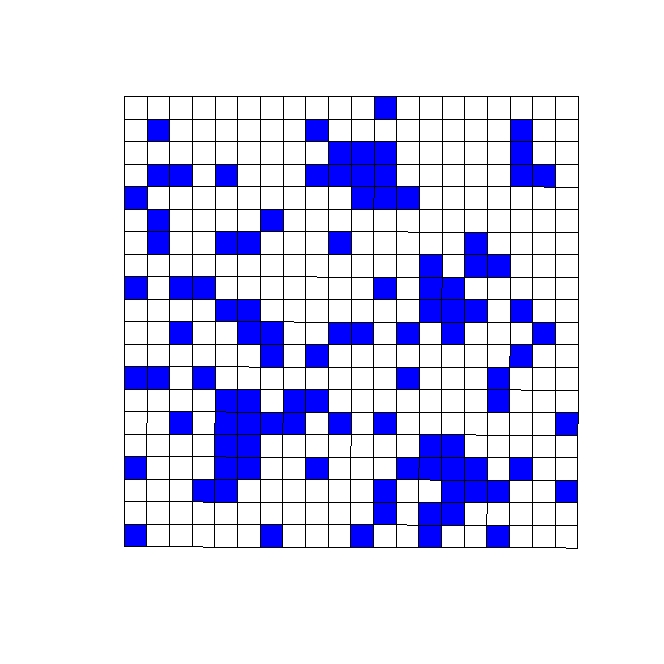}
\includegraphics[scale = 0.11,trim={3cm 3cm 3cm 3cm},clip]{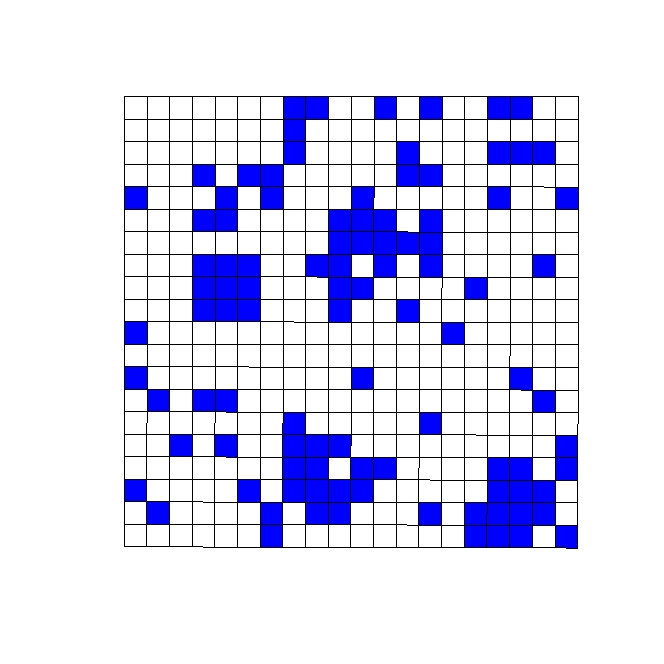}
\includegraphics[scale = 0.11,trim={3cm 3cm 3cm 3cm},clip]{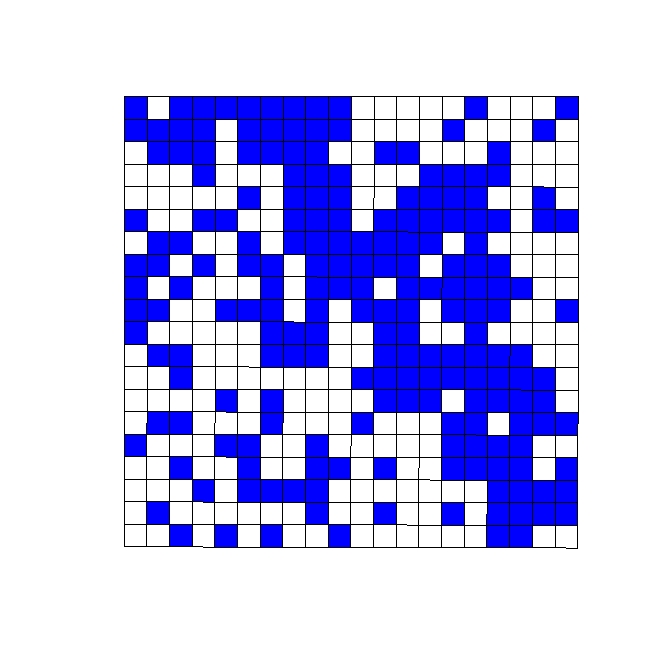}
\includegraphics[scale = 0.11,trim={3cm 3cm 3cm 3cm},clip]{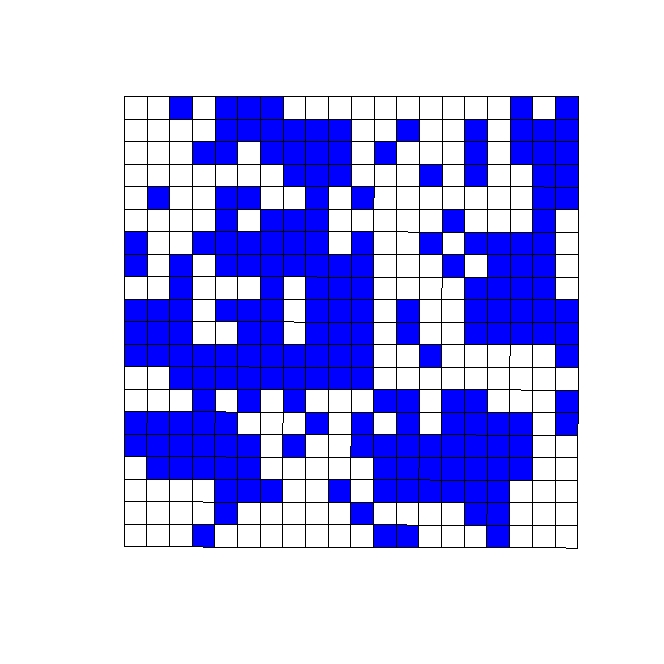}
}
\centerline{
\includegraphics[scale = 0.11,trim={3cm 3cm 3cm 3cm},clip]{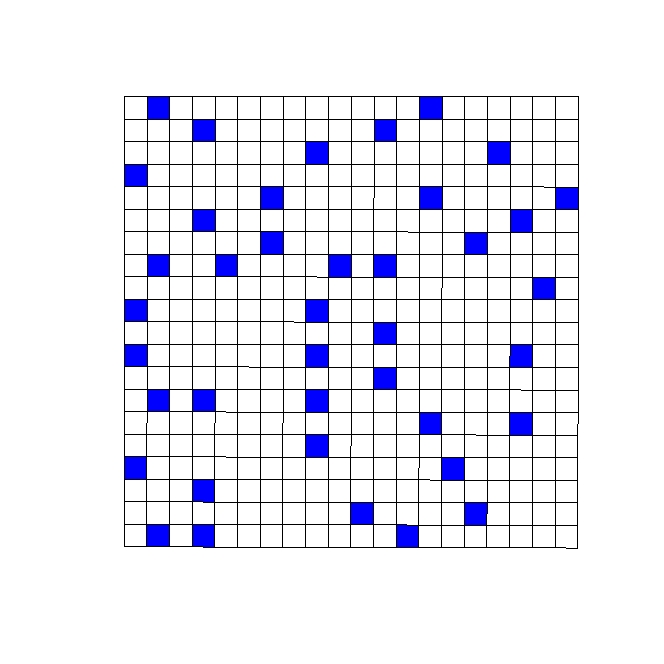}
\includegraphics[scale = 0.11,trim={3cm 3cm 3cm 3cm},clip]{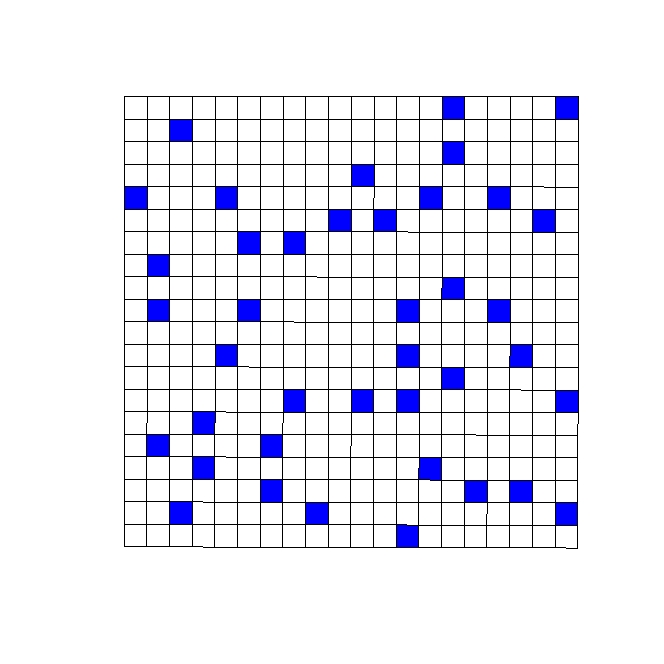}
\includegraphics[scale = 0.11,trim={3cm 3cm 3cm 3cm},clip]{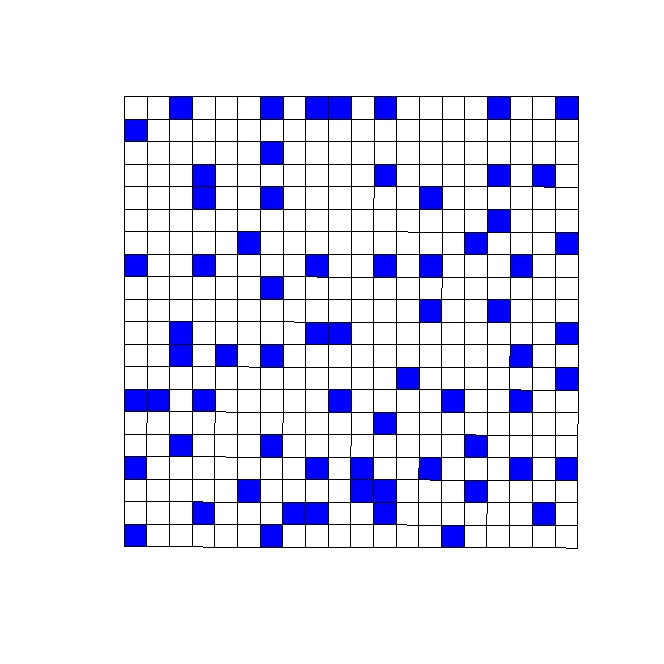}
\includegraphics[scale = 0.11,trim={3cm 3cm 3cm 3cm},clip]{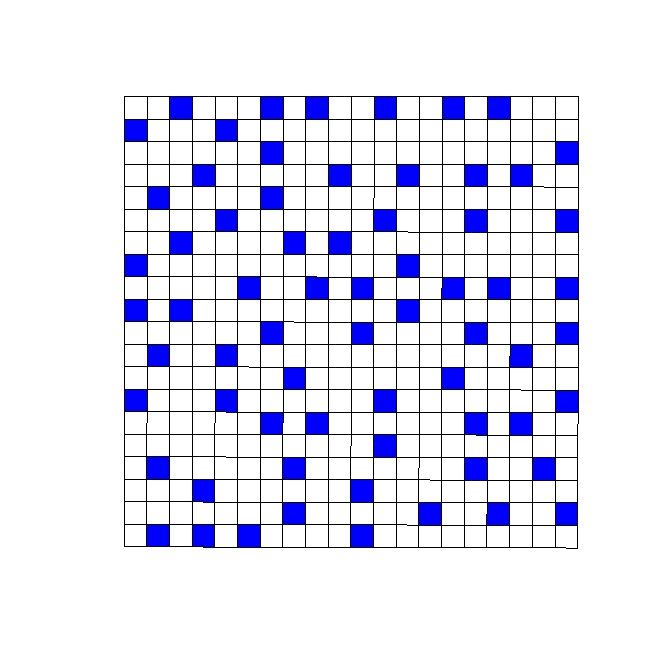}
}
\caption{Examples of observed units generated under each scenario for study area $\mathscr A_1$.  The first row displays examples of data generated under the null hypothesis of equal probability sampling without replacement (left to right: $I_1$, $I_2$, $I_3$).  The second row displays examples of data generated with a single cluster (left to right $C_1$, $C_2$, $C_3$, $C_4$, $C_5$, $C_6)$.  The third row displays examples of data generated with multiple clusters (left to right $C_7$, $C_8$, $C_9$, $C_{10}$, $C_{11}$, $C_{12}$).  The fourth row displays examples of data generated with dispersion (left to right $D_{1}$, $D_{2}$, $D_{3}$, $D_{4}$).\label{fig:gridsims}}
\end{figure}

Finally, we assess the ability of our hypothesis testing procedure to detect spatial dispersion under the following 4 scenarios.
\begin{align*}
        \textbf{$D_1$: }& \bs Y \sim C\left(\lceil N_j/10 \rceil,\lceil N_j/100 \rceil,\frac{1}{10}\right)\\
    \textbf{$D_2$: }& \bs Y \sim C\left(\lceil N_j/10 \rceil,\lceil N_j/100 \rceil, 0\right)\\
    \textbf{$D_3$: }& \bs Y \sim C\left(\lceil N_j/6 \rceil,\lceil N_j/100 \rceil,\frac{1}{10}\right)\\
    \textbf{$D_4$: }& \bs Y \sim C\left(\lceil N_j/6 \rceil,\lceil N_j/100 \rceil, 0\right)
\end{align*}
Here, in $D_{1}$ and $D_3$ adjacent units are one tenth as likely to be observed as non-adjacent units, creating a mild dispersion effect.  Under $D_{2}$ and $D_4$ adjacent units cannot be selected at all, creating a stronger dispersion effect.  Finally, we consider only two samples sizes when assessing dispersion ($N_j/10$ and $N_j/6$) because it becomes increasingly difficult or impossible to select only non-adjacent units as number of selected units increases.  Examples of data generated each scenario for $\mathscr A_1$ is shown in Figure \ref{fig:gridsims}.  Supplementary Figure 1 provides similar examples for $\mathscr A_2$.

\subsection{Method Specifications}

For each data generation scenario, we compare four methods:  the average nearest neighbor method applied to the areal unit centriods, the spatial scan statistic under a binomial likelihood with each areal unit is assumed to have population 1 and zone membership is determined using areal unit centriods, traditional Ripley's K function with Ripley's isotropic edge correction \citep{Ripley88}, and our RKAD.  For each study area, 10 radii $r_1,\ldots,r_{10}$ were selected for evaluation of the spatial scan statistic,  Ripley's K, and RKAD.  The set of radii used for each study area are shown in Figure \ref{fig:simsunits}.  The smallest radius was equal the smallest distance between any areal unit centriods, and the radii were increased incrementally, with the largest radii being approximately one fourth of the width of the study area.

The average nearest neighbor test statistic was computed using the \texttt{nni} function in the \texttt{spatialEco} R package \citep{spatialEco-package}, and the spatial scan statistic was computed using the \texttt{scan.test} function in the \texttt{spatstat} R package \citep{spatstat}.  Ripley's K function was computed using the \texttt{Kest} function in the R \texttt{spatstat} package.  The \texttt{envelope} function (also in the \texttt{spatstat} package) with $100$ simulations was used to perform hypothesis testing for Ripley's K function.

To perform the hypothesis testing procedure based on Ripley's K function for areal data, Monte Carlo simulations were used.  For each study area $\mathscr A_j$, $j = 1,2$ and each sample size $n \in \{ \lceil N_j/10 \rceil,\lceil N_j/6 \rceil, \lceil N_j/4 \rceil,\lceil N_j/2 \rceil \} $, 1,000 datasets were simulated under the null hypothesis of equal probability sampling without replacement.  For each study area $\mathscr A_j$ and $n$, $\bs y_g$ was generated from SWoR($N_j, n, \bs p_j)$ for $g =1,...,1000$, and 
\[
\widehat m(r_l; n) =  \frac{1}{1000}\sum_{g=1}^{1000} \left( \frac{1}{n}\sum_{i: y_{gi} = 1}  \frac{|c_i(r_l) \cap \mathscr B(\bs y_g) |}{\pi r_l^2} \right)
\]
was calculated.  To approximate the null distribution of the test statistic $T$ for each study area $\mathscr A_j$, observation size $n$, and radius $r_l$
\[
T_g(r_l,\bs y_g) = \frac{1}{n}\sum_{i: y_{gi} = 1} |m(r_l,i, \bs y_g) - \widehat m(r_l;n)|
\]
was calculated, and the quantiles of the set $\{ T_1(r_l,\bs y_1),...,T_{1000}(r_l, \bs y_{1000}) \}$ were used as approximate critical values for the hypothesis test.

After approximating the each null distribution, 500 instances of $\bs y$ were generated under each of the 19 data generating mechanisms.  For each radius $r_l$ and each $\bs y$, $T(r_l, \bs y)$ was computed and compared to critical values from the null distribution with the same observation size $n$ and radius $r_l$.  For data generation mechanisms $I_1$, $I_2$ and $I_3$, an $\alpha = 0.05$ two-tailed test was performed for each method.  Note that the spatial scan statistic method, the Ripley's K method, and the Ripley's K for areal data method all require Monte Carlo approximation of the null distribution, rendering the level of these tests approximate.  For data generation mechanisms $C_1-C_{12}$, a one-tailed test was performed with clustering as the alternative hypothesis.  Finally, for data generation methods $D_{1}-D_{4}$, a one tailed test was performed with dispersion as the alternative hypothesis.

\subsection{Simulation Results}

Tables \ref{tab:A1Sims} and \ref{tab:A2Sims} summarize the results from study area $\mathscr A_1$ (the regular grid) and study area $\mathscr A_2$ (the US counties), respectively.  Each table reports the empirical rate of rejection for the null hypothesis.  For scenarios $I_1$, $I_2$ and $I_3$, this quantity is the empirical type I error rate; for the other scenarios, this quantity is the empirical power.  As the Ripley's K method and the RKAD were performed at 10 different radii, the rejection rate for each radius is reported separately.

\begin{table}[b]
\caption{Simulation study results for study area $\mathscr A_1$ (the regular grid).  Results displayed include the empirical rejection rate of the average nearest neighbor method (ANN, column 2), the spatial scan statistic method (SST, column 3), the Ripley's K method at each of the 10 radii (RK, columns 5-14) and the Ripley's K for areal data method at each fo the 10 radii (RKAD, columns 5-14).  For data generation mechanisms (DGMs) $I_1$, $I_2$, and $I_3$, the reported rejection rates correspond to the type I error of an $\alpha \approx 0.05$ two-tailed test.  For DGMs $C_1-C_{12}$, the rejection rates correspond to the power of an $\alpha \approx 0.05$ single-tailed test indicative of clustering.  For DGMs $D_{1}-D_{4}$, the rejection rates correspond to the power of an $\alpha \approx 0.05$ single-tailed test indicative of dispersion.}
\tiny{
    \centering
       \begin{tabular}{lllllllllllllll}
    \hline 
    \hline
    \multirow{2}{*}{DGM} &
    \multirow{2}{*}{ANN} &\multirow{2}{*}{SST} &\multirow{2}{*}{Method} &  \multicolumn{10}{c}{Radius}  \\
    \cline{9-10}
    &&&&$r_1$ & $r_2$ & $r_3 $& $r_4$ & $r_5$ & $r_6$ & $r_7$ &$ r_8 $& $r_9$ & $r_{10}$\\
    \hline
   \multirow{2}{*}{$I_1$}& \multirow{2}{*}{97.5} & \multirow{2}{*}{3.0}& RK & 67.8 & 12.6 &30.0&23.0 &19.2 & 6.4 &19.0&  9.8&  6.4& 26.2 \\
&&&RKAD & 4.2 &3.8& 4.8 &3.6& 3.4& 4.4 &4.6 &4.6& 4.2& 4.2 \\
    \multirow{2}{*}{$I_2$}& \multirow{2}{*}{100.0}& \multirow{2}{*}{3.0}&RK & 100.0 & 63.4 & 96.6 & 90.4 & 78.6  & 4.4 & 77.4 & 24.4 & 13.6 & 82.0 \\
    &&&RKAD & 4.8 &  6.4 & 6.4 & 6.0 & 5.4 & 3.8 & 4.8 & 4.8 & 3.0 & 4.0\\
    \multirow{2}{*}{$I_3$} & \multirow{2}{*}{100.0} & \multirow{2}{*}{2.8} & RK &100.0 & 100.0 &100.0 &100.0& 100.0  & 0.6& 100.0 & 75.8 & 40.4& 100.0 \\
    &&&RKAD &7.6& 6.4 &5.8 &5.0 &4.6 &3.8& 5.0& 5.2 &5.6& 4.6 \\
    \cline{1-14}
    \multirow{2}{*}{$C_1$}&\multirow{2}{*}{96.5}&\multirow{2}{*}{84.8} &RK & 0.0 & 82.6& 14.4& 94.8& 51.6& 78.4 &68.6 &84.4 &87.0& 77.0\\
    &&& RKAD & 36.0 &42.8 &62.0& 69.4 &78.0 &85.0& 89.2 &91.8 &91.6& 93.0 \\
    \multirow{2}{*}{$C_2$}&\multirow{2}{*}{95.8}&\multirow{2}{*}{99.8}& RK &  0.0 & 98.8 & 66.2& 100.0 & 94.6 & 99.4 & 99.2 & 99.8& 100.0 & 99.4 \\
    &&& RKAD & 76.4 & 85.8 & 95.6 & 97.6 &100.0 & 99.8& 100.0& 100.0& 100.0& 100.0 \\
    \multirow{2}{*}{$C_3$}&\multirow{2}{*}{100.0}&\multirow{2}{*}{100.0}& RK &  0.0&  100.0&   14.8&  100.0 &  84.8 &  99.4 &  93.8&   98.8 &  99.4 &  95.6 \\
    &&& RKAD & 97.6  & 98.4 &  99.8 &  99.8 &  99.8 &  99.8 & 100.0 & 100.0 & 100.0 & 100.0\\
        \multirow{2}{*}{$C_4$}&\multirow{2}{*}{100.0}&\multirow{2}{*}{100.0}& RK & 0.0&  100.0&   97.2&  100.0&  100.0&  100.0 & 100.0 & 100.0&  100.0&  100.0 \\
    &&& RKAD &  100.0 & 100.0& 100.0& 100.0& 100.0 &100.0 &100.0 &100.0& 100.0 &100.0\\
    \multirow{2}{*}{$C_5$}&\multirow{2}{*}{100.0}&\multirow{2}{*}{100.0}& RK &  0.0 & 100.0 &  0.0& 100.0 &  9.4 & 99.8  &54.4&  99.6 &100.0 & 69.8\\
    &&& RKAD & 100.0& 100.0 &100.0& 100.0 &100.0 &100.0 &100.0 &100.0& 100.0& 100.0\\
        \multirow{2}{*}{$C_6$}& \multirow{2}{*}{100.0}& \multirow{2}{*}{100.0} &RK & 0.0& 100.0&   0.0 &100.0 & 90.6 &100.0 &100.0 &100.0 &100.0 &100.0 \\
    &&& RKAD &100.0 &100.0 &100.0 &100.0& 100.0 &100.0 &100.0 &100.0 &100.0 &100.0 \\
    \hline
    \multirow{2}{*}{$C_7$}& \multirow{2}{*}{88.4} & \multirow{2}{*}{36.4} & RK & 0.2 &82.6 &14.6 &75.2& 18.2 &28.4& 10.2& 13.0& 13.2&  4.6  \\
    &&& RKAD & 38.8 &46.0 &52.8& 39.6& 35.2& 30.6 &25.8 &21.0 &17.2 &14.2 \\
        \multirow{2}{*}{$C_8$}&\multirow{2}{*}{86.2}&\multirow{2}{*}{78.0} & RK & 0.4& 99.4 &65.4& 98.0 &67.0& 72.2& 45.4& 45.0 &42.2& 20.6 \\
    &&& RKAD &87.2 &90.8 &93.8 &89.6 &83.8 &75.4& 63.4 &52.2 &42.4& 37.0 \\
    \multirow{2}{*}{$C_9$}&\multirow{2}{*}{100.0}&\multirow{2}{*}{60.2}& RK & 0.0& 100.0  & 4.0 &100.0&  13.6 & 63.2 &  7.0 & 17.4  &20.2&   2.2 \\
    &&& RKAD &95.4& 97.2& 96.2& 90.8& 81.8& 72.0& 64.8& 54.4& 44.2& 40.2 \\
        \multirow{2}{*}{$C_{10}$}&\multirow{2}{*}{100.0}&\multirow{2}{*}{95.4}&  RK & 0.0 & 100.0 &  60.2 &  100.0 &  78.6 &  95.0 &  45.4 &  58.6 &  57.6 &  19.6 \\
    &&& RKAD & 100.0 &  100.0 &  100.0 & 100.0 &  99.4 &  97.2 &  92.4 &  85.6 &  74.8 &  67.8\\
    \multirow{2}{*}{$C_{11}$}& \multirow{2}{*}{100.0}& \multirow{2}{*}{74.1} &RK &0.0 &100.0  & 0.0 &100.0  & 0.0 & 61.2 &  0.0 &  3.0  & 6.8  & 0.0 \\
    &&& RKAD &99.2& 100.0& 100.0  &98.0 & 93.2 & 86.4 & 76.6 & 69.0 & 61.0 & 58.0 \\
        \multirow{2}{*}{$C_{12}$}& \multirow{2}{*}{100.0}& \multirow{2}{*}{93.6} &RK &  0.0 &100.0  & 0.0& 100.0  & 1.6 & 92.0  & 1.8 & 20.8 & 28.6  & 0.4 \\
    &&& RKAD &100.0& 100.0& 100.0& 100.0&  99.6 & 97.8 & 95.2 &  90.2 & 85.2  &80.4 \\
    \hline
\multirow{2}{*}{$D_{1}$}&\multirow{2}{*}{0.0} &\multirow{2}{*}{1.0} & RK & 98.0 & 95.8& 100.0 &  3.4 & 83.6 & 26.8 & 67.8 & 32.6 & 16.8&  58.8 \\
    &&& RKAD &17.0&  0.2 & 0.4 & 5.0 & 5.8 &17.2 &19.4& 19.6 &22.8& 22.0 \\
        \multirow{2}{*}{$D_{2}$}& \multirow{2}{*}{0.0} & \multirow{2}{*}{1.0}& RK & 100.0 &100.0& 100.0 &  9.0 & 91.6&  30.6 & 79.4 & 37.2 & 21.6  &65.4 \\
    &&& RKAD & 15.2 & 0.0 & 0.0 & 3.0 & 5.2 &23.6 &27.2 &21.8 &27.6 &29.2\\
    \multirow{2}{*}{$D_{3}$}&\multirow{2}{*}{0.0}&\multirow{2}{*}{1.2} & RK & 100.0 & 99.6 &100.0 &  0.2 & 98.8 & 35.4 & 98.6 & 66.0 & 46.0 & 97.0\\
    &&& RKAD & 0.4 & 0.0 & 0.0 & 8.2 &36.4 &50.6& 55.8& 57.6 &57.6 &54.0 \\
    \multirow{2}{*}{$D_{4}$}&\multirow{2}{*}{0.0}&\multirow{2}{*}{3.8}& RK &100.0 &100.0& 100.0&   0.0 & 99.8 & 50.2 &100.0 & 88.2 & 66.8 & 99.8 \\
    &&& RKAD & 0.0 & 0.0 & 0.0 & 6.2& 71.4 &92.2 &96.6& 93.8 &88.4& 86.8\\
    \hline \hline
    \end{tabular}
    }
    \label{tab:A1Sims}
\end{table}

\begin{table}[t]
\caption{Simulation study results for study area $\mathscr A_2$ (the US counties).  Results displayed include the empirical rejection rate of the average nearest neighbor method (ANN, column 2), the spatial scan statistic method (SST, column 3), the Ripley's K method at each of the 10 radii (RK, columns 5-14) and the Ripley's K for areal data method at each fo the 10 radii (RKAD, columns 5-14).  For data generation mechanisms (DGMs) $I_1$, $I_2$, and $I_3$, the reported rejection rates correspond to the type I error of an $\alpha \approx 0.05$ two-tailed test.  For DGMs $C_1-C_{12}$, the rejection rates correspond to the power of an $\alpha \approx 0.05$ single-tailed test indicative of clustering.  For DGMs $D_{1}-D_{4}$, the rejection rates correspond to the power of an $\alpha \approx 0.05$ single-tailed test indicative of dispersion.}
\tiny{
    \centering
       \begin{tabular}{lllllllllllllll}
    \hline 
    \hline
    \multirow{2}{*}{DGM} &
    \multirow{2}{*}{ANN} & \multirow{2}{*}{SST}&\multirow{2}{*}{Method} &  \multicolumn{10}{c}{Radius}  \\
    \cline{9-10}
    &&&&$r_1$ & $r_2$ &$ r_3$ & $r_4$ & $r_5$ & $r_6$ & $r_7$ & $r_8$ & $r_9$ & $r_{10}$\\
    \hline
\multirow{2}{*}{$I_1$}& \multirow{2}{*}{26.8} & \multirow{2}{*}{3.6} & RK & 0.6 & 99.4& 100.0& 100.0 &100.0 &100.0 &100.0 &100.0 &100.0 &100.0\\
&&&RKAD & 7.4 &5.4& 5.0 &4.0 &4.2 &4.0& 4.2 &4.6 &4.4 &4.8 \\
    \multirow{2}{*}{$I_2$}& \multirow{2}{*}{10.0} &\multirow{2}{*}{4.2}&  RK & 95.8 &100.0& 100.0 &100.0 &100.0 &100.0& 100.0& 100.0 &100.0& 100.0 \\
    &&&RKAD &3.8 &4.8& 3.6& 3.8 &3.4 &3.4& 4.0 &3.8& 3.4& 3.8\\
    \multirow{2}{*}{$I_3$}&\multirow{2}{*}{100.0}&\multirow{2}{*}{2.8} & RK & 0.0 & 100.0 &100.0& 100.0& 100.0 &100.0 &100.0& 100.0& 100.0& 100.0 \\
    &&&RKAD & 5.8 &2.8 &4.2& 4.6& 5.0& 4.0& 5.0& 5.0 &5.4 &6.0\\
    \cline{1-14}
\multirow{2}{*}{$C_1$} & \multirow{2}{*}{0} &\multirow{2}{*}{100.0} &  RK &80.4 &100.0& 100.0& 100.0&100.0 &100.0& 100.0 &100.0 &100.0& 100.0 \\
    &&& RKAD & 8.2 & 61.2& 100.0& 100.0 &100.0 &100.0 &100.0& 100.0 &100.0 &100.0  \\
    \multirow{2}{*}{$C_2$}&\multirow{2}{*}{94.0}&\multirow{2}{*}{100.0}& RK & 3.6 &100.0& 100.0& 100.0& 100.0& 100.0& 100.0& 100.0 &100.0& 100.0  \\
    &&& RKAD & 9.2 & 61.2 &100.0& 100.0 &100.0& 100.0 &100.0 &100.0& 100.0& 100.0  \\
    \multirow{2}{*}{$C_3$}&\multirow{2}{*}{0}&\multirow{2}{*}{100.0}& RK & 0.6 &100.0 &100.0 &100.0 &100.0 &100.0& 100.0& 100.0 &100.0 &100.0   \\
    &&& RKAD &  2.8 &100.0& 100.0 &100.0 &100.0 &100.0 &100.0& 100.0 &100.0 &100.0\\
        \multirow{2}{*}{$C_4$}&\multirow{2}{*}{0.0} &\multirow{2}{*}{100.0} & RK & 0.8 &100.0 &100.0 &100.0 &100.0 &100.0& 100.0 &100.0 &100.0 &100.0 \\
    &&& RKAD & 3.2 &100.0 &100.0& 100.0& 100.0& 100.0&100.0& 100.0& 100.0 &100.0 \\
    \multirow{2}{*}{$C_5$}&\multirow{2}{*}{99.6}& \multirow{2}{*}{100.0}& RK & 0.0 &100.0& 100.0& 100.0 &100.0 &100.0 &100.0 &100.0& 100.0 &100.0 \\
    &&& RKAD & 0.2& 100.0 &100.0& 100.0& 100.0& 100.0& 100.0& 100.0& 100.0& 100.0\\
        \multirow{2}{*}{$C_6$}& \multirow{2}{*}{96.4} & \multirow{2}{*}{100.0} &RK &  0.0 & 100.0 &100.0 &100.0 &100.0 &100.0 &100.0 &100.0& 100.0& 100.0 \\
    &&& RKAD &0.6& 100.0 &100.0& 100.0 &100.0 &100.0& 100.0& 100.0 &100.0 &100.0\\
    \hline
    \multirow{2}{*}{$C_7$}& \multirow{2}{*}{0} & \multirow{2}{*}{47.4}& RK &  77.2 &100.0& 100.0& 100.0& 100.0 &100.0 &100.0 &100.0 &100.0 &100.0  \\
    &&& RKAD & 3.8 &68.4 &44.4 &30.8 &25.4& 21.8 &18.4& 15.4& 12.6 &13.2\\
        \multirow{2}{*}{$C_8$}&\multirow{2}{*}{0.2} &\multirow{2}{*}{87.8}  & RK & 78.6 &100.0& 100.0& 100.0 &100.0& 100.0 &100.0 &100.0& 100.0 &100.0\\
    &&& RKAD & 2.2& 99.2& 89.0& 71.4& 55.8& 44.4& 36.0 &32.6& 27.8& 25.4\\
    \multirow{2}{*}{$C_9$}&\multirow{2}{*}{0.0} &\multirow{2}{*}{78.4} & RK & 0.0& 100.0 &100.0 &100.0& 100.0 &100.0 &100.0 &100.0& 100.0& 100.0\\
    &&& RKAD &  2.4 &100.0 & 92.6 & 77.2 & 63.4 & 49.4 & 39.0 & 29.4 & 25.8 & 23.4\\
        \multirow{2}{*}{$C_{10}$}&\multirow{2}{*}{0.0} &\multirow{2}{*}{98.6} &RK& 0.0 & 100.0 &100.0 &100.0 &100.0& 100.0& 100.0& 100.0 &100.0 &100.0 \\
    &&& RKAD & 0.8 & 100.0 & 99.6 & 97.8 & 91.4 & 82.6 & 71.0 & 61.2 & 52.2  &47.4 \\
    \multirow{2}{*}{$C_{11}$}& \multirow{2}{*}{100.0}  & \multirow{2}{*}{100.0}&RK & 0.0 & 100.0& 100.0& 100.0& 100.0& 100.0& 100.0& 100.0 &100.0& 100.0 \\
    &&& RKAD & 2.8& 100.0 & 99.4 & 96.2 & 87.6 & 74.4 & 59.2 & 48.2 & 41.6 & 35.0\\
        \multirow{2}{*}{$C_{12}$}& \multirow{2}{*}{100.0} & \multirow{2}{*}{98.6}&RK & 0.0& 100.0& 100.0 &100.0 &100.0 &100.0 &100.0& 100.0 &100.0 &100.0  \\
    &&& RKAD & 2.8 &100.0 &100.0 & 99.8&  98.4  &91.8&  84.0&  73.8  &67.4 & 58.0
 \\
    \hline
\multirow{2}{*}{$D_{1}$}&\multirow{2}{*}{0.0} &\multirow{2}{*}{0.0}& RK & 100.0 &  0.8  & 0.0 &  0.0  & 0.0 &  0.0 &  0.0 &  0.0  & 0.0 &  0.0
\\
    &&& RKAD & 11.6 &32.2& 63.4& 51.2& 43.4 &33.4& 28.6& 28.8 &25.0 &22.4\\
        \multirow{2}{*}{$D_{2}$}& \multirow{2}{*}{0.0}& \multirow{2}{*}{0.0} & RK & 100.0 &  4.6  & 0.0  & 0.0  & 0.0 &  0.0  & 0.0  & 0.0  & 0.0 &  0.0\\
    &&& RKAD & 8.4 &39.2 &79.6 &68.0& 58.0& 43.6& 35.0 &32.6 &26.6& 23.8\\
    \multirow{2}{*}{$D_{3}$}&\multirow{2}{*}{0.0} &\multirow{2}{*}{0.0}& RK &99.0 & 0.0 & 0.0  &0.0 & 0.0 & 0.0 & 0.0  &0.0  &0.0 & 0.0 \\
    &&& RKAD& 2.2& 67.6 &97.0& 87.0& 71.8& 58.2 &46.8 &36.4& 26.6& 24.4  \\
    \multirow{2}{*}{$D_{4}$}&\multirow{2}{*}{0.0}&\multirow{2}{*}{0.0}& RK &100.0 &  0.0  & 0.0  & 0.0 &  0.0 &  0.0 &  0.0  & 0.0 &  0.0 &  0.0\\
    &&& RKAD &1.0 & 80.6& 100.0&  97.0&  86.0 & 70.4  &56.8  &36.2 & 25.8 & 21.8\\
    \hline \hline
    \end{tabular}
    }
    \label{tab:A2Sims}
\end{table}

Under the null scenarios ($I_1$, $I_2$ and $I_3$), the empirical type I error rate of RKAD is within a Monte Carlo margin of error of its nominal level for all radii considered.  The empirical type I error rate of the spatial scan statistic is at (or slightly below) its nominal level for these scenarios.  The empirical type I error rate of the point process methods were markedly inflated, rejecting 100\% of the null tests in some cases. The ANN method was generally worse for the regular grid than for the US counties, while Ripley's K was generally worse for the US counties and situations with more observations.

Under the large-scale clustering scenarios ($C_1-C_6$), the RKAD method has high empirical power to detect clustering for all radii except the smallest radius, with performance improving as the strength of the clustering and sample size increase.  The empirical power of the spatial scan statistic method is also quite high.  Under the small-scale clustering scenarios ($C_7-C_{12})$, the empirical power of the RKAD method is generally high $(> 85\%)$ for radii $r_2-r_3$, with power declining somewhat for larger radii.  As the clustering in these scenarios occurs at a close geographic scale (i.e. first degree neighbors), we would expect the highest power at the smaller radii.  The exception is scenario $C_7$, which corresponds to the weakest clustering and the smallest sample size, for which empirical power is notably lower. The empirical power of RKAD at radii $r_2$ and $r_3$ is greater than or equal to that of the spatial scan statistic for almost all multiple clustering scenarios. 

Under the dispersion scenarios ($D_{1}$- $D_{4}$), the empirical power of RKAD varies considerably by scenario.  For the regular grid, performance is rather poor for the smaller sample size, but improves as the strength of the dispersion and the number of observed units increases.  Performance is much better for the US counties.  The empirical power of the spatial scan statistic is quite low ($<5\%$) for all dispersion scenarios.

In summary, of the four methods considered, only RKAD and the spatial scan statistic had satisfactory type I error rates.  As expected, the empirical power of RKAD varied with the choice of radius, with empirical power being higher for radii which correspond to the type of clustering present in the data (i.e. larger radii for the large-scale clustering in scenarios $C_1-C_6$ and smaller radii for the small-scale clustering in scenarios $C_7-C_{12}$).  In almost all scenarios, the empirical power of RKAD with the ideal radius was higher than that of the spatial scan statistic.

\section{Data Application}

In this Section, we consider the performance of our method on real world applications from two different fields.  First, we use the method to determine if land parcels held as CEs are clustered in Boulder County, Colorado, using the 112,819 distinct land parcels in Boulder County as the areal structure.  Next, we apply our method to determine if US counties with high childhood overweight rates are spatially clustered, using the 3,108 county and county-equivalents in the contiguous US as the areal structure.

\subsection{Application to Conservation Easements}

CEs are a private and generally perpetual form of land conservation that legally severs aspects of private landownership (e.g., development rights, resource extraction, etc.) from a parcel of land \citep{McLaughlin09}. Although a landowner makes an individual decision to place a CE, there is evidence of spatial clustering of CEs over time, throughout the US \citep{Lamichhane21}. Cumulative and clustered CE use may impact regional ecosystem character by altering the degree of CE parcels' isolation or connectivity with other ecologically valuable parcels and may change ecologic quality on the CE parcel itself \citep{Graves19}. The greater the mass of clustering and ecological systems integrity, the more impact there may be on the land conversion rates at the county level, and on the decision to leave a parcel in open space (or not), potentially affecting placement of other socially valuable land uses as well.  Furthermore, recognizing if and where CEs are clustered and linking the social, political, biological, and geographical characteristics to the clustered areas may help elucidate the factors driving CE placement \citep{Baldwin15}.

\begin{figure}
    \centering
    \includegraphics[scale = 0.3,trim={1cm 0cm 2cm 4cm},clip,angle = 350]{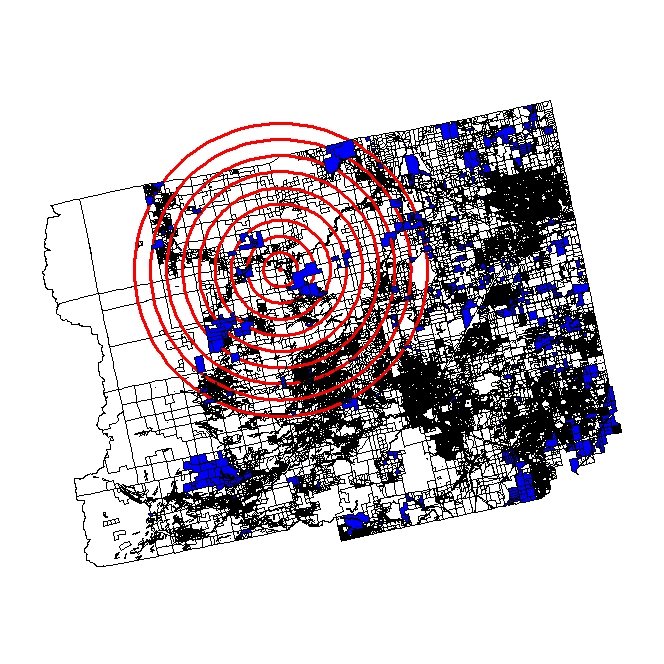}
    \includegraphics[scale = 0.4,trim={3cm 0cm 2cm 4cm},clip,angle = 345]{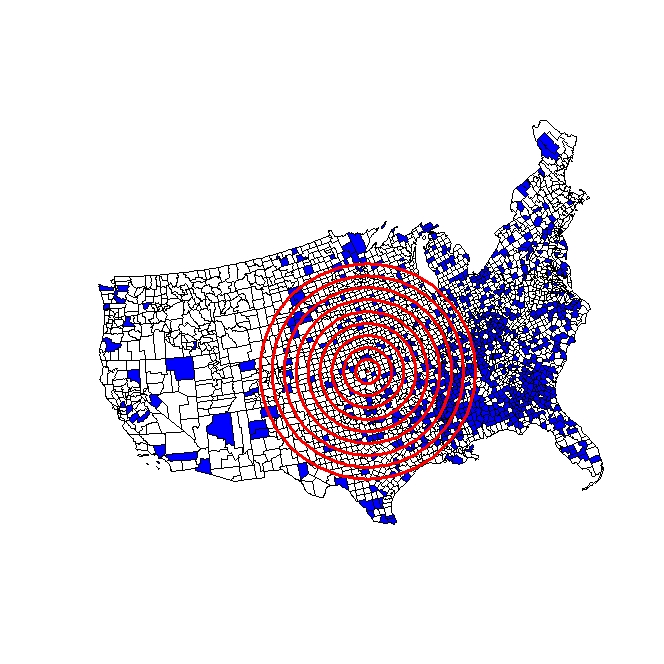}
  \vspace{-0.8in}
    \caption{The top pane displays the 112,819 land parcels in Boulder County, Colorado in 2008.  Parcels held as a CE are shown in blue.  The radii at which the RKAD method was evaluated are shown in red.  Note that areas with many small parcels appear black.  The bottom pane displays the 3,108 counties in the contiguous US.  Counties with a high rate of childhood overweight/obesity are shown in blue.  The radii at which the RKAD method was evaluated are shown in red.}
    \label{fig:DataApp}
\end{figure}

The Boulder County data consists of 112,819 land parcels in place in 2008.  Of these land parcels, 817 were held as CEs.  A parcel was considered to be part of a CE if any part of the parcel was part of an easement.  Figure \ref{fig:DataApp} depicts the land parcels; parcels which are part of CE are shown in blue.  The method was applied at 10 different radii, also depicted in Figure \ref{fig:DataApp}.

In order to apply our method, the distribution of the RKAD test statistic under the null hypothesis was estimated using 1,000 Monte Carlo simulations.  In each Monte Carlo simulation, 817 parcels were selected via simple random sampling without replacement.  The observed RKAD test statistic was larger than the 95th quantile of the estimated null distribution for all radii, indicating that parcels which contain CEs are significantly clustered for all radii.

Human land conversion has caused widespread habitat loss and fragmentation \citep{Haddad15}. In the context of CEs with a purpose of biological conservation, clustering easements close to one another is one reserve design principle to improve landscape connectivity and combat the adverse effects of habitat fragmentation \citep{Diamond75}. Larger and higher quality habitats (easements) increase the size and stability of source populations and subsequently increase species dispersal capabilities \citep{Hodgson09}. Clustering and structural connectivity between conservation areas are not always positive, however, as clustering may also leave these areas vulnerable to spatially autocorrelated extinction pressures, such as diseases, invasive species, stochastic environmental events, or negative effects from localized urban growth \citep{Donaldson16}. Given that RKAD indicated spatial clustering of CEs in Boulder County, more detailed landscape connectivity studies focused on functional connectivity may be warranted \citep{Balbi19,Tischendorf00}. 

\subsection{Application to Counties with High Childhood Overweight/Obesity Rates}

Next, we use the RKAD method to determine if US counties with high childhood overweight/obesity rates are spatially clustered.  County-level childhood overweight rates were estimated from data collected in the 2016 National Survey of Children's Health using a multilevel small area estimation approach as described in \cite{Zgodic21}. A county was considered to have a high overweight rate if its estimated rate exceeded the 75th percentile of all county overweight rates.  There are 3,108 counties and county-equivalents in the contiguous US, and 786 of these counties were found to have a high rate of childhood overweight.  These counties are shown in blue in Figure \ref{fig:DataApp}, along with the radii at which RKAD was applied.

The distribution of the RKAD test statistic under the null hypothesis was estimated using 1,000 Monte Carlo simulations.  In each Monte Carlo simulation, 786 counties were selected via simple random sampling without replacement.  The observed RKAD test statistic was larger than the 95th quantile of the estimated null distribution for all radii, indicating that counties with high rates of childhood overweight are significantly clustered.  As Southeastern and Midwest states tend to have higher overweight and obesity rates than the rest of the country \citep{Gartner16, CDC16}, these results are not surprising.

\section{Conclusion}

The problem of assessing areal data for spatial clustering has been the subject of relatively little attention.  While the ANN method and the traditional Ripley's K method are often used to assess areal data for clustering by mapping each areal unit to its centroid, these methods were not designed for areal data.  Our simulation study shows that applying these methods in this manner results in a highly inflated type I error rate. In fact, in many settings these methods rejected 100\% of the null hypotheses. Since such an approach is the default method used by ArcGIS software, these results are concerning.  Among the existing methods assessed in our simulation study, only the spatial scan statistic maintained its nominal type I error rate.

To address the relative lack of methods to assess areal data for clustering, we developed RKAD, an extension of Ripley's K function which is capable of assessing areal data for the presence of clustering or dispersion at specific geographic scales.  The RKAD method quantifies the average amount of observed area within a specified distance of each areal unit centroid.  RKAD can be used to perform a hypothesis test for the presence of spatial clustering or dispersion by comparing the observed RKAD test statistic to the distribution of the RKAD test statistic in the absence of spatial dependence. Simulation studies demonstrated that RKAD hypothesis testing procedure maintains its nominal type I error rate and has high power to detect a variety of spatial patterns, including small and large scale clustering and dispersion.  RKAD generally displayed higher empirical power than the spatial scan statistic, especially when data were dispersed.

To facilitate the use of our method, R code which implements the RKAD method and performs the necessary Monte Carlo simulations has been made available online at https://github.com/scwatson812/RKAD.  When the number of observed areal units is large, the Monte Carlo simulations may be run in parallel to reduce computation time.  The development of faster methods for approximating the null distribution is an excellent area for future work.


\section*{Funding}
SS and AM were supported in part by the Research Center for Child Well-Being [NIGMS P20GM130420].  SS, AZ, and AM were supported in part by the Centers for Disease Control [5 U19 DD 001218].  SS, AO, DW, and CD were supported in part by the National Science Foundation [CNH-L 1518455].  The funding sources played no role in study design, data collection, data analysis, or manuscript publication.

\section*{Competing Interests}
Declarations of interest: none

\section*{Supplementary Material}
The Web Appendix contains contains Supplementary Figure 1.


\section*{References}
\bibliographystyle{apa}

\begin{thebibliography}{}

\bibitem[\protect\astroncite{Amgad et~al.}{2015}]{Amgad15}
Amgad, M., Itoh, A., and Tsui, M. M.~K. (2015).
\newblock Extending ripley's k-function to quantify aggregation in 2-d
  grayscale images.
\newblock {\em PLOS ONE}, 10(12):1--22.


\bibitem[\protect\astroncite{Andersen}{1992}]{Andersen92}
Andersen, M. (1992).
\newblock Spatial analysis of two-species interaction.
\newblock {\em Oecologia}, 91:134--140.

\bibitem[\protect\astroncite{Baddeley et~al.}{2015}]{spatstat}
Baddeley, A., Rubak, E., and Turner, R. (2015).
\newblock {\em Spatial Point Patterns: Methodology and Applications with {R}}.
\newblock Chapman and Hall/CRC Press, London.

\bibitem[\protect\astroncite{Balbi et~al.}{2019}]{Balbi19}
Balbi, M., Petit, E.~J., Croci, S., Nabucet, J., Georges, R., Madec, L., and
  Ernoult, A. (2019).
\newblock Title: Ecological relevance of least cost path analysis: An easy
  implementation method for landscape urban planning.
\newblock {\em Journal of Environmental Management}, 244:61--68.

\bibitem[\protect\astroncite{Baldwin and Leonard}{2015}]{Baldwin15}
Baldwin, R. and Leonard, P. (2015).
\newblock Interacting social and environmental predictors for the spatial
  distribution of conservation lands.
\newblock {\em PLOS ONE}, 10(10).

\bibitem[\protect\astroncite{Bhatt and Tiwari}{2016}]{Bhatt16}
Bhatt, V. and Tiwari, N. (2016).
\newblock A spatial scan statistic for survival data based on generalized life
  distribution.
\newblock {\em Communications in Statistics - Theory and Methods},
  45(19):5730--5744.



\bibitem[\protect\astroncite{CDC}{2021}]{CDC16}
CDC (2021).
\newblock Trends and maps, https://nccd.cdc.gov/


\bibitem[\protect\astroncite{Clark}{1956}]{Clark56}
Clark, P.~J. (1956).
\newblock Grouping in spatial distributions.
\newblock {\em Science}, 123(3192):373--374.

\bibitem[\protect\astroncite{Clark and Evans}{1954}]{Clark54}
Clark, P.~J. and Evans, F.~C. (1954).
\newblock Distance to nearest neighbor as a measure of spatial relationships in
  populations.
\newblock {\em Ecology}, 35(4):445--453.


\bibitem[\protect\astroncite{Clark and Evans}{1955}]{Clark55}
Clark, P.~J. and Evans, F.~C. (1955).
\newblock On some aspects of spatial pattern in biological populations.
\newblock {\em Science}, 121(3142):397--398.

\bibitem[\protect\astroncite{Clark and Evans}{1979}]{Clark79}
Clark, P.~J. and Evans, F.~C. (1979).
\newblock Generalization of a nearest neighbor measure of dispersion for use in
  k dimensions.
\newblock {\em Ecology}, 60(2):316--317.

\bibitem[\protect\astroncite{Davarpanah et~al.}{2018}]{davarpanah2018spatial}
Davarpanah, A., Babaie, H.~A., and Dai, D. (2018).
\newblock Spatial autocorrelation of neogene-quaternary lava along the snake
  river plain, idaho, usa.
\newblock {\em Earth Science Informatics}, 11(1):59--75.

\bibitem[\protect\astroncite{de~Carvalho et~al.}{2021}]{Matos21}
de~Carvalho, D.~M., do~Amaral, G. J.~A., and Bastiani, F.~D. (2021).
\newblock Spatial scan statistics based on empirical likelihood.
\newblock {\em Communications in Statistics - Simulation and Computation},
  0(0):1--15.



\bibitem[\protect\astroncite{Diamond}{1975}]{Diamond75}
Diamond, J. (1975).
\newblock The island dilemma: lessons of modern biogeographic studies for the
  design of natural reserves.
\newblock {\em Biological Conservation}, 7:129--146.

\bibitem[\protect\astroncite{Diggle}{1983}]{Diggle83}
Diggle, P. (1983).
\newblock {\em Statistical analysis of spatial point patterns}.
\newblock Academic Press, London.

\bibitem[\protect\astroncite{Dixon}{2014}]{Dixon14}
Dixon, P.~M. (2014).
\newblock {\em Ripley's K Function}.
\newblock American Cancer Society.

\bibitem[\protect\astroncite{Donaldson et~al.}{2016}]{Donaldson16}
Donaldson, L., Wilson, R., and Maclean, I. (2016).
\newblock Old concepts, new challenges: adapting landscape-scale conservation
  to the twenty-first century.
\newblock {\em Biodiversity and Conservation}, 26(3):527--552.

\bibitem[\protect\astroncite{ESRI}{2021a}]{ArcANN}
ESRI (2021a).
\newblock Average nearest neighbor.

\bibitem[\protect\astroncite{ESRI}{2021b}]{arcRipK}
ESRI (2021b).
\newblock Multi-distance spatial cluster analysis (ripley's k function)
  (spatial statistics).

\bibitem[\protect\astroncite{Evans}{2021}]{spatialEco-package}
Evans, J.~S. (2021).
\newblock {\em spatialEco}.
\newblock R package version 1.3-6.

\bibitem[\protect\astroncite{Gartner et~al.}{2016}]{Gartner16}
Gartner, D.~R., Taber, D.~R., Hirsch, J.~A., and Robinson, W.~R. (2016).
\newblock The spatial distribution of gender differences in obesity prevalence
  differs from overall obesity prevalence among us adults.
\newblock {\em Annals of Epidemiology}, 26(4):293--298.

\bibitem[\protect\astroncite{Getis and Franklin}{1987}]{Getis87}
Getis, A. and Franklin, J. (1987).
\newblock Second-order neighborhood analysis of mapped point patterns.
\newblock {\em Ecology}, 68(3):473--477.


\bibitem[\protect\astroncite{Goreaud and P\'elissier}{1999}]{Goreaud99}
Goreaud, F. and P\'elissier, R.~e. (1999).
\newblock On explicit formulas of edge effect correction for ripley's
  k-function.
\newblock {\em Journal of Vegetation Science}, 10(3):433--438.

\bibitem[\protect\astroncite{Graves et~al.}{2019}]{Graves19}
Graves, R., Williamson, M., Belote, T., and Brandt, J. (2019).
\newblock Quantifying the contribution of conservation easements to
  large-landscape conservation.
\newblock {\em Biological Conservation}, 232:83--96.

\bibitem[\protect\astroncite{Haase}{1995}]{Haase95}
Haase, P. (1995).
\newblock Spatial pattern analysis in ecology based on ripley's k-function:
  Introduction and methods of edge correction.
\newblock {\em Journal of Vegetation Science}, 6(4):575--582.

\bibitem[\protect\astroncite{Haddad et~al.}{2015}]{Haddad15}
Haddad, N.~M., Brudvig, L.~A., Clobert, J., Davies, K.~F., Gonzalez, A., Holt,
  R.~D., Lovejoy, T.~E., Sexton, J.~O., Austin, M.~P., Collins, C.~D., Cook,
  W.~M., Damschen, E.~I., Ewers, R.~M., Foster, B.~L., Jenkins, C.~N., King,
  A.~J., Laurance, W.~F., Levey, D.~J., Margules, C.~R., Melbourne, B.~A.,
  Nicholls, A.~O., Orrock, J.~L., Song, D.-X., and Townshend, J.~R. (2015).
\newblock Habitat fragmentation and its lasting impact on earth's ecosystems.
\newblock {\em Science Advances}, 1(2):e1500052.

\bibitem[\protect\astroncite{Hodgson et~al.}{2009}]{Hodgson09}
Hodgson, J.~A., Thomas, C.~D., Wintle, B.~A., and Moilanen, A. (2009).
\newblock Climate change, connectivity and conservation decision making: back
  to basics.
\newblock {\em Journal of Applied Ecology}, 46(5):964--969.

\bibitem[\protect\astroncite{Huang et~al.}{2007}]{Huang07}
Huang, L., Kulldorff, M., and Gregorio, D. (2007).
\newblock A spatial scan statistic for survival data.
\newblock {\em Biometrics}, 63(1):109--118.

\bibitem[\protect\astroncite{Huang et~al.}{2009}]{Huang09}
Huang, L., Tiwari, R.~C., Zou, Z., Kulldorff, M., and Feuer, E.~J. (2009).
\newblock Weighted normal spatial scan statistic for heterogeneous population
  data.
\newblock {\em Journal of the American Statistical Association},
  104(487):886--898.

\bibitem[\protect\astroncite{Jung et~al.}{2007}]{Jung07}
Jung, I., Kulldorff, M., and Klassen, A.~C. (2007).
\newblock A spatial scan statistic for ordinal data.
\newblock {\em Statistics in Medicine}, 26(7):1594--1607.

\bibitem[\protect\astroncite{Karunaweera et~al.}{2020}]{Karunaweera2020}
Karunaweera, N.~D., Ginige, S., Senanayake, S., Silva, H., Manamperi, N.,
  Samaranayake, N., Siriwardana, Y., Gamage, D., Senerath, U., and Zhou, G.
  (2020).
\newblock Spatial epidemiologic trends and hotspots of leishmaniasis, sri
  lanka, 2001-2018.
\newblock {\em Emerging infectious diseases}, 26.

\bibitem[\protect\astroncite{Kretser et~al.}{2008}]{Kretser}
Kretser, H., Sullivan, P., and Knuth, B. (2008).
\newblock Housing density as an indicator of spatial patterns of reported
  human-wildlife interactions in northern new york.
\newblock {\em Landscape and Urban Planning}, 84:282--292.

\bibitem[\protect\astroncite{Kulldorff}{1997}]{Kulldorff97}
Kulldorff, M. (1997).
\newblock A spatial scan statistic.
\newblock {\em Communications in Statistics - Theory and Methods},
  26(6):1481--1496.

\bibitem[\protect\astroncite{Kulldorff et~al.}{2006}]{Kulldorff06}
Kulldorff, M., Huang, L., Pickle, L., and Duczmal, L. (2006).
\newblock An elliptic spatial scan statistic.
\newblock {\em Statistics in Medicine}, 25(22):3929--3943.

\bibitem[\protect\astroncite{Lamichhane et~al.}{2021}]{Lamichhane21}
Lamichhane, S., Sun, C., Gordon, J., Grado, S., and Poudel, K. (2021).
\newblock Spatial dependence and determinants of conservation easement
  adoptions in the united states.
\newblock {\em Journal of Environmental Management}, 296.

\bibitem[\protect\astroncite{Lee and Lee}{2013}]{Lee13}
Lee, S.-K. and Lee, B. (2013).
\newblock Assessing the appropriateness of the spatial distribution of standard
  lots using the l-index.
\newblock {\em Journal of the Korean Society of Surveying, Geodesy,
  Photogrammetry and Cartography}, 31(6.2):601--609.

\bibitem[\protect\astroncite{Marj and Abellan}{2013}]{Tonini13}
Marj, T. and Abellan, A. (2013).
\newblock Rockfall detection from terrestrial lidar point clouds: A clustering
  approach using r.
\newblock {\em Journal of Spatial Information Science}, 8.

\bibitem[\protect\astroncite{Martins et~al.}{2009}]{Martins09}
Martins, L., Silva, A., Paiva, A., and Gattass, M. (2009).
\newblock Detection of breast masses in mammogram images using growing neural
  gas algorithm and ripley's k function.
\newblock {\em Journal of Signal Processing Systems - JSPS}, 55:77--90.

\bibitem[\protect\astroncite{McLaughlin and Weeks}{2009}]{McLaughlin09}
McLaughlin, N. and Weeks, W. (2009).
\newblock In defense of conservation easements: A response to the end of
  perpetuity.
\newblock {\em Wyoming Law Review}, 9:1--96.

\bibitem[\protect\astroncite{Philo and Philo}{2021}]{Philo21}
Philo, C. and Philo, P. (2021).
\newblock 2.15 or not 2.15? an historical-analytical inquiry into the
  nearest-neighbor statistic.
\newblock {\em Geographical Analysis}.

\bibitem[\protect\astroncite{Qiao et~al.}{2019}]{Qiao19}
Qiao, L., Huang, H., and Tian, Y. (2019).
\newblock The identification and use efficiency evaluation of urban industrial
  land based on multi-source data.
\newblock {\em Sustainability}, 11(21).

\bibitem[\protect\astroncite{Ripley}{1981}]{Ripley81}
Ripley, B. (1981).
\newblock {\em Spatial Statistics}.
\newblock Wiley, New York, NY.

\bibitem[\protect\astroncite{Ripley}{1988}]{Ripley88}
Ripley, B. (1988).
\newblock {\em Statistical Inference for Spatial Processes}.
\newblock Cambridge University Press.

\bibitem[\protect\astroncite{Ripley}{1976}]{Ripley76}
Ripley, B.~D. (1976).
\newblock The second-order analysis of stationary point processes.
\newblock {\em Journal of Applied Probability}, 13(2):255--266.

\bibitem[\protect\astroncite{Ripley}{1977}]{Ripley77}
Ripley, B.~D. (1977).
\newblock Modelling spatial patterns.
\newblock {\em Journal of the Royal Statistical Society. Series B
  (Methodological)}, 39(2):172--212.

\bibitem[\protect\astroncite{Sayer and Wienhold}{2013}]{Sayer13}
Sayer, D. and Wienhold, M. (2013).
\newblock A gis-investigation of four early anglo-saxon cemeteries: Ripley's
  k-function analysis of spatial groupings amongst graves.
\newblock {\em Social Science Computer Review}, 31(1):71--89.

\bibitem[\protect\astroncite{Shen and Jiang}{2014}]{Shen14}
Shen, X. and Jiang, W. (2014).
\newblock Multivariate normal spatial scan statistic for detecting the most
  severe cluster of a disease.
\newblock {\em Journal of Management Analytics}, 1(2):130--145.

\bibitem[\protect\astroncite{Siordia}{2013}]{Siordia13}
Siordia, C. (2013).
\newblock Benefits of small area measurements: a spatial clustering analysis on
  medicare beneficiaries in the usa.
\newblock {\em Human Geographies - Journal of Studies and Research in Human
  Geography}, 7(`):53--59.

\bibitem[\protect\astroncite{Skog et~al.}{2014}]{Skog2014}
Skog, L., Linde, A., Palmgren, H., Hauska, H., and Elgh, F. (2014).
\newblock Spatiotemporal characteristics of pandemic influenza.
\newblock {\em BMC Infectious Diseases}, 14.

\bibitem[\protect\astroncite{Soltani and Aboukhamseen}{2015}]{Soltani15}
Soltani, A.~R. and Aboukhamseen, S.~M. (2015).
\newblock An alternative cluster detection test in spatial scan statistics.
\newblock {\em Communications in Statistics - Theory and Methods},
  44(8):1592--1601.

\bibitem[\protect\astroncite{Sterner et~al.}{1986}]{Sterner86}
Sterner, R.~W., Ribic, C.~A., and Schatz, G.~E. (1986).
\newblock Testing for life historical changes in spatial patterns of four
  tropical tree species.
\newblock {\em Journal of Ecology}, 74(3):621--633.

\bibitem[\protect\astroncite{Szwagrzyk and Czerwczak}{1993}]{Szwagrzyk93}
Szwagrzyk, J. and Czerwczak, M. (1993).
\newblock Spatial patterns of trees in natural forests of east-central europe.
\newblock {\em Journal of Vegetation Science}, 4(4):469--476.

\bibitem[\protect\astroncite{Tango and Takahashi}{2005}]{Tango2005}
Tango, T. and Takahashi, K. (2005).
\newblock {A flexibly shaped spatial scan statistic for detecting clusters}.
\newblock {\em International Journal of Health Geographics}, 4(1):11.

\bibitem[\protect\astroncite{Tischendorf and Fahrig}{2000}]{Tischendorf00}
Tischendorf, L. and Fahrig, L. (2000).
\newblock On the usage and measurement of landscape connectivity.
\newblock {\em Oikos}, 90(1):7--19.

\bibitem[\protect\astroncite{Upton and Fingleton}{1985}]{Upton85}
Upton, G. and Fingleton, B. (1985).
\newblock {\em Spatial Data Analysis byExample. Vol. 1. Point Pattern and
  Quantitative Data}.
\newblock John Wiley, New York, NY.

\bibitem[\protect\astroncite{Wade}{2014}]{Wade14}
Wade, B.~J. (2014).
\newblock Spatial analysis of global prevalence of multiple sclerosis suggests
  need for an updated prevalence scale.
\newblock {\em Multiple Sclerosis International}, 2014.

\bibitem[\protect\astroncite{{Yunta} et~al.}{2014}]{Yunta14}
{Yunta}, M.~L., {Lagache}, T., {Santi-Rocca}, J., {Bastin}, P., and
  {Olivo-Marin}, J. (2014).
\newblock A statistical analysis of spatial clustering along cell filaments
  using ripley's k function.
\newblock In {\em 2014 IEEE 11th International Symposium on Biomedical Imaging
  (ISBI)}, pages 541--544.

\bibitem[\protect\astroncite{Zgodic et~al.}{2021}]{Zgodic21}
Zgodic, A., Eberth, J.~M., Breneman, C.~B., Wende, M.~E., Kaczynski, A.~T.,
  Liese, A.~D., and McLain, A.~C. (2021).
\newblock {Estimates of Childhood Overweight and Obesity at the Region, State,
  and County Levels: A Multilevel Small-Area Estimation Approach}.
\newblock {\em American Journal of Epidemiology}.
\newblock kwab176.

\bibitem[\protect\astroncite{Zipp et~al.}{2017}]{Zipp17}
Zipp, K.~Y., Lewis, D.~J., and Provencher, B. (2017).
\newblock Does the conservation of land reduce development? an
  econometric-based landscape simulation with land market feedbacks.
\newblock {\em Journal of Environmental Economics and Management}, 81:19 -- 37.



\end{thebibliography}

\end{document}